\documentclass[fleqn,usenatbib]{mnras}

\usepackage{newtxtext,newtxmath}

\usepackage{multicol}        % Multi-column entries in tables
\usepackage{bm}		% Bold maths symbols, including upright Greek
\usepackage{pdflscape}	% Landscape pages
\usepackage{tikz}
\usepackage{soul}
\usepackage{caption}
\usepackage{graphicx} % Required for including pictures
\usepackage{wrapfig} % Allows in-line images
\usepackage{amsmath}
\usepackage{physics}
\usepackage{subcaption}
\usepackage{rotating}
\usepackage{url}

\graphicspath{{figures/}}
\newcommand{\ms}{\mbox{\,m s$^{-1}$}}
\newcommand{\cms}{\mbox{\,cm s$^{-1}$}}
\usepackage[T1]{fontenc}

\DeclareRobustCommand{\VAN}[3]{#2}
\let\VANthebibliography\thebibliography
\def\thebibliography{\DeclareRobustCommand{\VAN}[3]{##3}\VANthebibliography}

\usepackage{graphicx}	% Including figure files
\usepackage{amsmath}	% Advanced maths commands
\title[Astrometric bias in barycentric correction]{Mitigating astrometric bias in barycentric correction with PEXO}

\author[Yicheng Rui and Fabo Feng]{
Yicheng Rui,$^{1,2}$
Fabo Feng$^{1,2}$\thanks{Contact E-mail: \href{mailto:ffeng@sjtu.edu.cn}{ffeng@sjtu.edu.cn} (TDLI)}
\\
$^{1}$Tsung-Dao Lee Institute, Shanghai Jiao Tong University, Shengrong Road 520, Shanghai, 201210, People's Republic of China\\
$^{2}$School of Physics and Astronomy, Shanghai Jiao Tong University, 800 Dongchuan Road, Shanghai 200240, People's Republic of China
}

\date{Accepted XXX. Received YYY; in original form ZZZ}

\pubyear{2023}

\begin{document}
\onecolumn
\label{firstpage}
\pagerange{\pageref{firstpage}--\pageref{lastpage}}
\maketitle

\begin{abstract}
Extremely precise radial velocity is essential for the detection of sub-m\,s$^{-1}$ radial velocity of stars induced by Earth-like planets. Although modeling of the barycentric correction of radial velocity could achieve 1\,mm\,s$^{-1}$ precision, the input astrometry could be biased due to nonlinear motions of stars caused by companions. 
To account for astrometry-induced bias in barycentric correction, we correct for astrometric bias by minimizing the scatter of reduced RV data with PEXO. In particular, we apply this method to the barycentric correction for 266 stars from HARPS data archive. We find that the RV scatter for 8 targets are significantly reduced due to correction of astrometric bias. Among these targets, 2 targets exhibit bias caused by known massive companions, while for the remaining 6 targets, the bias could be attributed to  unknown companions or Gaia systematics. Furthermore, 14 targets have an astrometry-induced annual RV variation higher than 0.05 \ms, and 10 of them are closer than 10\,pc.  We show the results of Barnard's star as an example, and find that an annual RV bias of 10\,cm\,s$^{-1}$ is mitigated by replacing  {BarCor} by PEXO as the barycentric correction code. Our work demonstrates the necessity of astrometric bias correction {and the utilization of barycentric correction code within a  relativistic framework} in high-precision RV for the detection of Earth-like planets.
\end{abstract}


%%%%%%%%%%%%%%%%%%%%%%%%%%%%%%%%%%%%%%%%%%%%%%%%%%

%%%%%%%%%%%%%%%%% BODY OF PAPER %%%%%%%%%%%%%%%%%%

\section{Introduction} \label{sec:intro}
Inspired by the Large UV/Optical/Infrared Surveyor (LUVOIR) \citep{theluvoirteam2019luvoir} and the Habitable Exoplanet Observatory (HabEx) \citep{gaudi2020habitable}, the Habitable Worlds Observatory (HWO) is proposed to search for life on nearby Earth-like exoplanets \citep{Mamajek23}. One essential task to archive this goal is to detect several nearby Earth-like planets. 

Extremely precise radial velocity (EPRV) is important for the detection of small radial velocity (RV) signals caused by small planets. To detect the RV variation induced by an Earth-like planet (Earth-size planets in habitable zone), the noise caused by host star, instrument and data reduction pipeline should be reduced to a level of $\sim$0.1\ms. While RV facilities such as Echelle SPectrograph for Rocky Exoplanets and Stable Spectroscopic Observations (ESPRESSO)  mounted on the European Southern Observatory's Very Large Telescope (VLT) \citep{pepe21} can achieve an instrumental prediction of about 0.1\ms, the data reduction pipeline and stellar activity introduce the main challenges to further improvement of long-term RV precision for the detection of Earth-like planets. 

Many sources of stellar noise should be mitigated in order to reliably detecting Earth-like planets. For example, star spots and plages would induce RV signals of $>1$\ms \citep{star_activity}; Star convection have the effect of  $\sim 500$\ms for solar-type stars \citep{solarconvection}; Effects of micro-telluric lines on precise radial velocities would lead to a noise of $>0.1$\ms \citep{telluric}.
Advanced noise modeling of time-correlated noise (e.g. \citealt{Feng2016,rajpaul2017}), line-by-line analysis by \citealt{Dumusque_2018} and \citealt{Lisogorskyi2019} as well as high-cadence RV measurements \citep{Hall_2018} are proposed to disentangle planetary signals from stellar noises. %The improvement of instrumental precision and modeling approach is demonstrated by the recent detection of 0.29\ms RV signals in the ESPRESSO RV data \citep{Barros2022}. 

Data reduction puts another challenge to EPRV. One of the main components in RV data reduction is barycentric correction, which subtracts the motion of the Earth from the measured RVs of a star to derive the RV variation of the star viewed from the heliocenter. This so-called   ``barycentric correction'' has been well modeled to 1\,$\mathrm{mm\ s}^{-1}$ precision in recently years \citep{Wright14,Feng19}. However, these models require precise input astrometry, precise mid-exposure time, and solar system ephemerides. The first requirement is not satisfied when a star move nonlinearly due to companion perturbations, while the other two requirements could be easily satisfied by using color-dependent high-precision photon counters \citep{pepe21,Jurgenson16} and well updated JPL ephemerides \citep{Folkner14}. 

The nonlinear motion of a star (or reflex motion) caused by unknown companions would vary the line-of-sight direction and the projection of Earth's velocity along this direction. Because both the nonlinear stellar motion and the Earth's annual motion would change the line of sight, the biased astrometry inferred under the assumption of linear stellar motion (e.g. \citealt{Gaia2016,Gaia2018,Gaia2021,GaiaDR3}) would not only introduce an RV variation with a period of the orbital motion of the unknown companion but also induce an annual RV variation caused by the coupling of the reflex motion and the Earth's motion. Considering that this coupled RV variation is quite similar to the RV signal induced by an Earth-like planet in terms of periodicity (300-400 days) and amplitude ($\sim$0.1\ms), it is crucial to mitigate the astrometric bias in barycentric correction. 

To mitigate the bias in input astrometry for barycentric correction, we model the barycentric RV with offsets in position and proper-motion. We use the Precise EXOplanetology (PEXO) package to calculate the gradient of RV relative to these parameters, and infer these parameters through likelihood maximization. We apply this approach to the High Accuracy Radial velocity Planet Searcher (HARPS) RV data (named ``RVbank'') reduced by \citealt{Trifonov20} using the SERVAL pipeline \citep{SERVAL}.
To date, this is the first time that astrometric bias is systematically considered in barycentric correction. 

This paper is structured as follows. In section \ref{sec:dataandtools}, we introduce the data we use in this work; in section \ref{sec:method}, we introduce the method for correction and testing; in section \ref{sec:results}, we show the main results of the correction; in section \ref{sec:cad}, we conclude and discuss the results.

\section{Data}\label{sec:dataandtools}
With a resolution of $R=115000$, HARPS \citep{HARPS} installed  at the ESO 3.6\,m telescope is one of the most productive high-resolution spectrographs in the detection of exoplanets using the radial velocity method. More than $6000$  stars have been observed by HARPS according to the ESO archive\footnote{\url{http://archive.eso.org/wdb/wdb/adp/phase3_main/form}}. Because the fiber of HARPS was changed in June 2015 \citep{HARPS_change_fiber}, we use the RV data obtained before June 1st, 2015 (JD=$2457174.5$) to avoid the potential RV offset caused by the change of fiber. %Because the fiber of HARPS was changed in June 2015 \citep{HARPS_change_fiber}, we considered the RV data obtained before  and after June 1st, 2015 (JD=$2457174.5$) independently to consider potential RV offset caused by fibre change. 

The raw spectra obtained by HARPS are reduced using the official data reduction software to deliver the reduced RV data \citep{HARPS}. The HARPS data collected before 2020 are re-reduced by \cite{Trifonov20} using the SpEctrum Radial Velocity AnaLyser (SERVAL) pipeline \citep{SERVAL}, which integrates the barycentric correction code {primarily based on BarCor \citep{Hrudkova09}} and routines specifically designed for various spectrographs. After removing the long-term drift caused by instrumental effects from the RV data, \cite{Trifonov20} obtain the RVbank HARPS data for $\sim 3000$ stars. {In SERVAL, BarCor is revised to consider proper motion of stars. Meanwhile, the secular acceleration (SA) is calculated separately. To make the results consistent with other pipelines, e.g. PEXO or Barycorr, which consider SA in barycentric correction, we compare the combined results of barycentric Earth radial velocity (BERV) and SA of SERVAL with PEXO or Barycorr.}

In \citealt{Trifonov20},    Simbad \citep{Simbad} astrometric data, which mainly comes from Gaia DR2, is used for the barycentric correction of the RVbank dataset. In this research, the input astrometric data used for most targets are from Gaia DR3. \citep{GaiaDR3}. For some bright targets without Gaia data, we use revised Hipparcos \citep{Leeuwen2007} data instead. The sources of reference astrometry are listed in our data table in the appendix section.

\begin{figure}
		\centering
		\includegraphics[width=1\columnwidth]{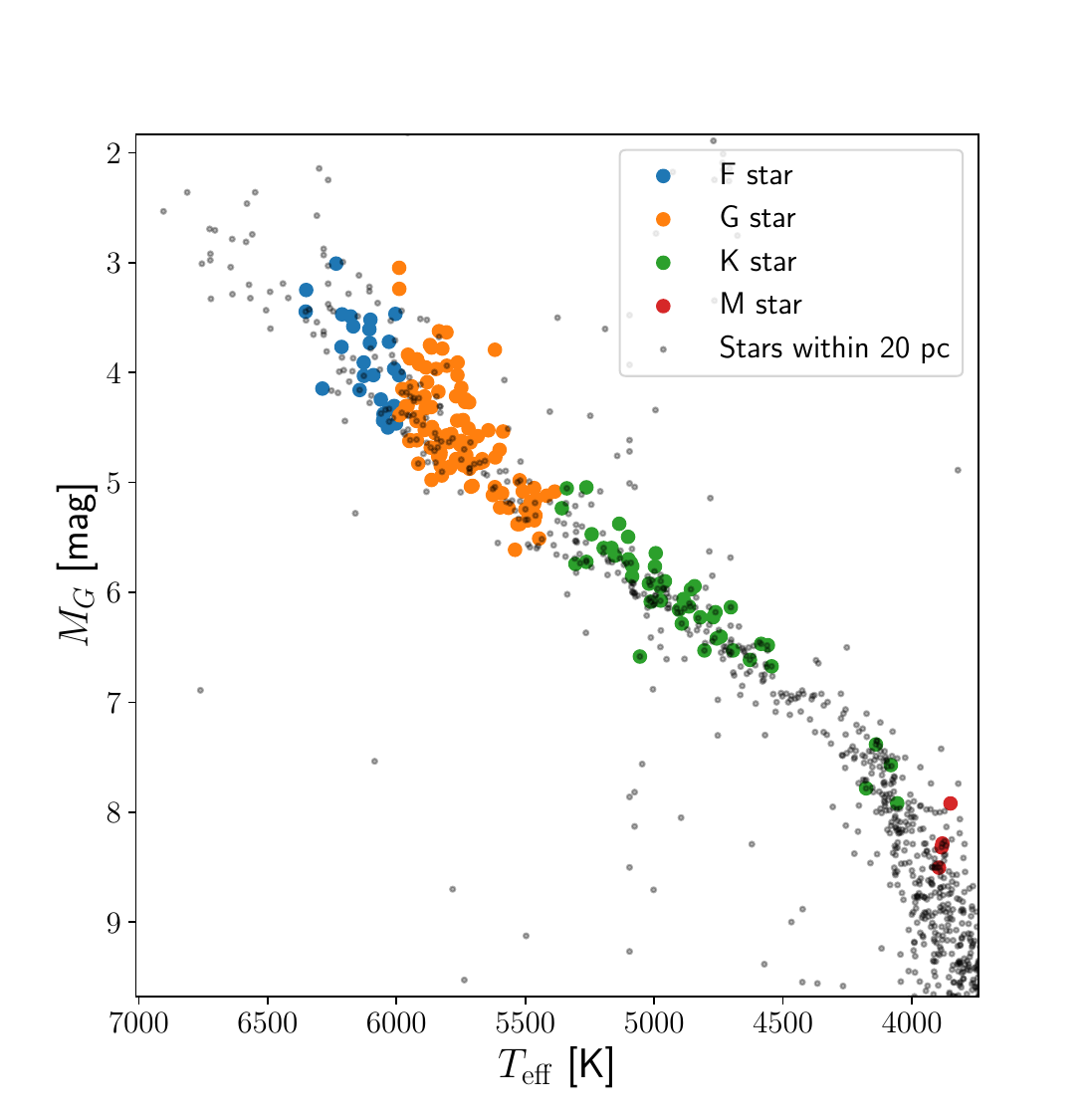}
		\caption{Hertzsprung–Russell diagram for selected targets. $T_{\rm eff}$ refers to the effective temperature from Gaia DR3; $M_G$ is the absolute magnitude derived from the Gaia DR3 magnitude and parallax. Color of the dots represents the type of the stars derived from $T_{\rm eff}$ \citep{Pecaut13}. Colored dots are the selected targets and grey dots are the Gaia DR3 targets closer than 20\,pc.}
		\label{fig:HRfig}
		\begin{flushleft}

		\end{flushleft}

	\end{figure}

Simbad \citep{Simbad} is used to crossmatch the RVbank targets with the Gaia and Hipparcos IDs. TOPCAT \citep{Taylor2005} is used to join the tables. We choose 266 targets according to the following criteria: 

\begin{enumerate}
    \item We remove the targets with RV variation $>30$\ms to avoid the contamination of binaries. Because we want to investigate the influence of wide companions on barycentric correction rather than the direct influence of binary companions on RV of the primary star; 
    \item The number of RV data points exceeds 50.  In order to  correct the sub-\ms RV variation induced by biased astrometry, we need a large amount of RV data with $\sim 1$\ms precision;
    \item The RV data is collected in more than 30 nights. This is to guarantee that the RV data is relatively evenly sampled such that the potential annual bias is regularly sampled;
    \item The observational time span for RV is longer than 100 days. This is because the potential astrometric bias caused by proper motion bias increases linearly over time, resulting in a significant bias for RV if the observational baseline is sufficiently long. %Additionally, the correction for annual signal would only be significant if the observational baseline is comparable or longer than one year.
\end{enumerate} %The crossmatch results for all targets which magnitude is bigger than $9$ are checked manually. 
The Hertzsprung–Russell diagram for these 266 targets is shown in Fig. \ref{fig:HRfig}. Most of the targets observed by HARPS are solar-type stars.  Additionally, the majority of these targets are located within a distance of 100 parsecs. The selected targets show a gap in the vicinity of 4500\, K, which is consistent with the distribution of nearby stars.%most of them in this sample have radial velocity measurements from Gaia. %HARPS only measures a subsample of the nearby Gaia main-sequence stars. More facilities are required to extend the current sample with EPRV data. 

\section{Method}\label{sec:method}

\subsection{Barycentric correction\label{sec:bc}}
To detect the perturbation of a star from its companion, the Earth's motion are typically subtracted from the observed RVs. It is often assumed that the target star moves linearly with respect to the Solar System barycenter. This barycentric correction is conducted in a relativistic framework \citep{Wright14,Feng19}. Current barycentric correction packages such as Barycorr \citep{Wright14} and PEXO \citep{Feng19} can model barycentric correction to a precision better than $1$\cms. They convert the Doppler shift of stellar spectrum measured on the Earth ($z_{\rm meas}$) to the Doppler shift that would be measured by a virtual observer at the barycenter of the solar system ($z_{\rm true}$) according to 
\begin{equation}
    1+z_{\rm true}=(1+z_B)(1+z_{\rm meas})~, 
    \label{eq:barycentriccorr}
    \end{equation}
where $z_B$ is the barycentric Doppler shift due to the Earth's motion around the solar system barycenter and various relativistic effects; $c$ is the speed of light. The radial velocity measured at the solar system barycenter is 
\begin{equation}
    v=c z_{\rm true}~. \label{eq:vr} 
\end{equation}

We conduct barycentric correction using PEXO, a global modeling framework for nanosecond timing, microsecond astrometry, and $\rm mm\,s^{-1}$ radial velocities \citep{Feng19}.  It is a global modeling framework which considers all the second-order effects including high-order Roemer delay and relativistic Doppler shift both in the solar system and in the target system. The philosophy of PEXO is to recommend simultaneous modeling of the barycentric motion of the observer and the reflex motion of the target to avoid bias caused by decoupling the Earth's motion and the reflex motion. Nevertheless, PEXO is suitable for high precision barycentric correction when the reflex motion is small or over long timescale. To optimize the astrometric offset, we use PEXO to numerically calculate the gradient of radial velocity with respect to the parameters of astrometric offsets. 

\subsection{Astrometric bias\label{sec:bias}}

\begin{figure}
    \centering
    	\begin{tikzpicture}
				
			% Define radius
			\def\r{3}
			\draw[dashed] (7,1) ellipse [x radius=1,y radius=1.2];
			% Bloch vector
			\draw (7, 1.66) node[circle, fill, inner sep=1.3, label=above:TSB] (a) {};
			
			\draw (0, 1.1) node[circle, fill, inner sep=1.9,label=above:E1] (e1) {};
			
			\draw (1, 0) node[circle, fill, inner sep=1.9,label=above:E2] (e2) {};
			
				\draw (0, -1.1) node[circle, fill, inner sep=1.9,,label=above:E3] (e3) {};
			\draw  (6, 1) node[circle, fill, inner sep=1.3, label=above:T] (b) {};
			
			\draw (0, 0.45) node[circle, fill, inner sep=2., label=above:Sun,color=orange] (s) {};
			%\draw[dashed] (orig) -- (\r/3, -\r/5) node (phi) {} -- (a);
			
			% Sphere
			\draw[dashed] (0,0) circle (1 and 1.1);
			\draw (b)++(-\r/5*3/5, -\r/5) node[circle, fill, inner sep=0.7, ] (ns1) {};
			\draw (b)++(-\r/5*3/5*2, -\r/5*2) node[circle, fill, inner sep=0.7, ] (ns2) {};
			\draw (e1)++(\r, \r/6*0.6) node[label=above:$\hat{n}$] (nbb1) {};
   \draw (e1)++(\r, -\r/5*0.11) node[label=below:$\hat{n}+\Delta\hat{n}$] (nbb2) {};
			\draw (a)++(-\r/5/5, -\r/5) node[circle, fill, inner sep=0.7, ] (nb1) {};
			\draw (a)++(-\r/5/5*2, -\r/5*2) node[circle, fill, inner sep=0.7, ] (nb2) {};
			
			% Axes
			\draw[->] (b) -- ++(-\r/5*3, -\r) node[below] (x1) {biased motion};
			
			%\draw (-\r/5*3/3, -\r/3) node(x11);
			%\draw (-\r/5*3/3*2, -\r/3*2) node(x12);

			\draw[->] (a) -- ++(-\r/5*1, -\r) node[below] (x2) {barycentric motion};
			
			\draw[->,red] (e1) -- (nbb1);
                \draw[dashed,red] (nbb1) -- (a);
			\draw[dashed,blue] (e2) -- (ns1);
			\draw[->,blue] (e1) -- (nbb2);
                \draw[dashed,blue] (nbb2) -- (b);
			\draw[dashed,red] (e2) --(nb1);
			\draw[dashed,blue] (e3) -- (ns2);
			\draw[dashed,red] (e3) --(nb2);
			%\draw[->] (s) -- ++(\r, -\r/5) node[below] (x3) {real $\vec{n}_r$};
			%\draw[->] (s) -- ++(\r*0.9, -\r/5*3*0.9) node[below] (x4) {observed $\vec{n}_r$};
			%\draw[->] (0,0) -- ++(0, \r) node[above] (x3) {$x_3$};
			
	\end{tikzpicture}
    
    \begin{flushleft}
    \caption{Effects of biased astrometry. E1, E2 and E3 represent the positions of the Earth at different epoch. $\hat{n}$ represents the direction from the Earth to the target system's barycenter (TSB). The biased target astrometry (T) provided by Gaia is expressed through five-parameter solutions. $\Delta \hat{n}$ represents the bias in the direction of the target's barycenter due to Gaia astrometry.\label{fig:model}}
    \end{flushleft}
\end{figure}
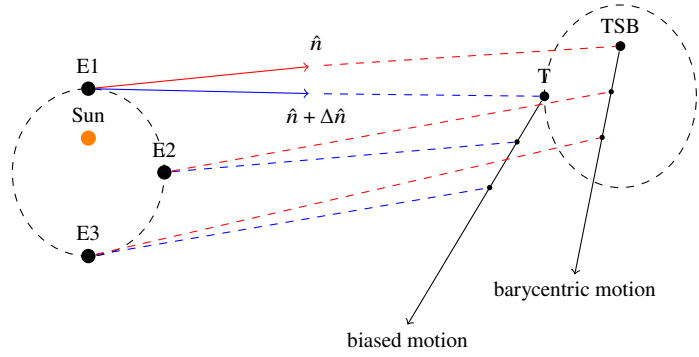

Unbiased barycentric correction is needed to obtain the precise motion of a target.  Fig. \ref{fig:model} illustrates the impact of biased astrometry on barycentric correction. Here we approximate the target motion by the motion of the target system's barycenter (TSB) of the target system because the positional difference between the target and the target system's barycenter does not introduce long-term RV variation or significant annual bias. Compared with the biased target astrometry given by Gaia, the barycentric motion has a much smaller astrometric offset with respect to the motion of the target and the offset does not accumulate with time.  

\begin{figure*}
    \centering
    \includegraphics[width=0.8\linewidth]{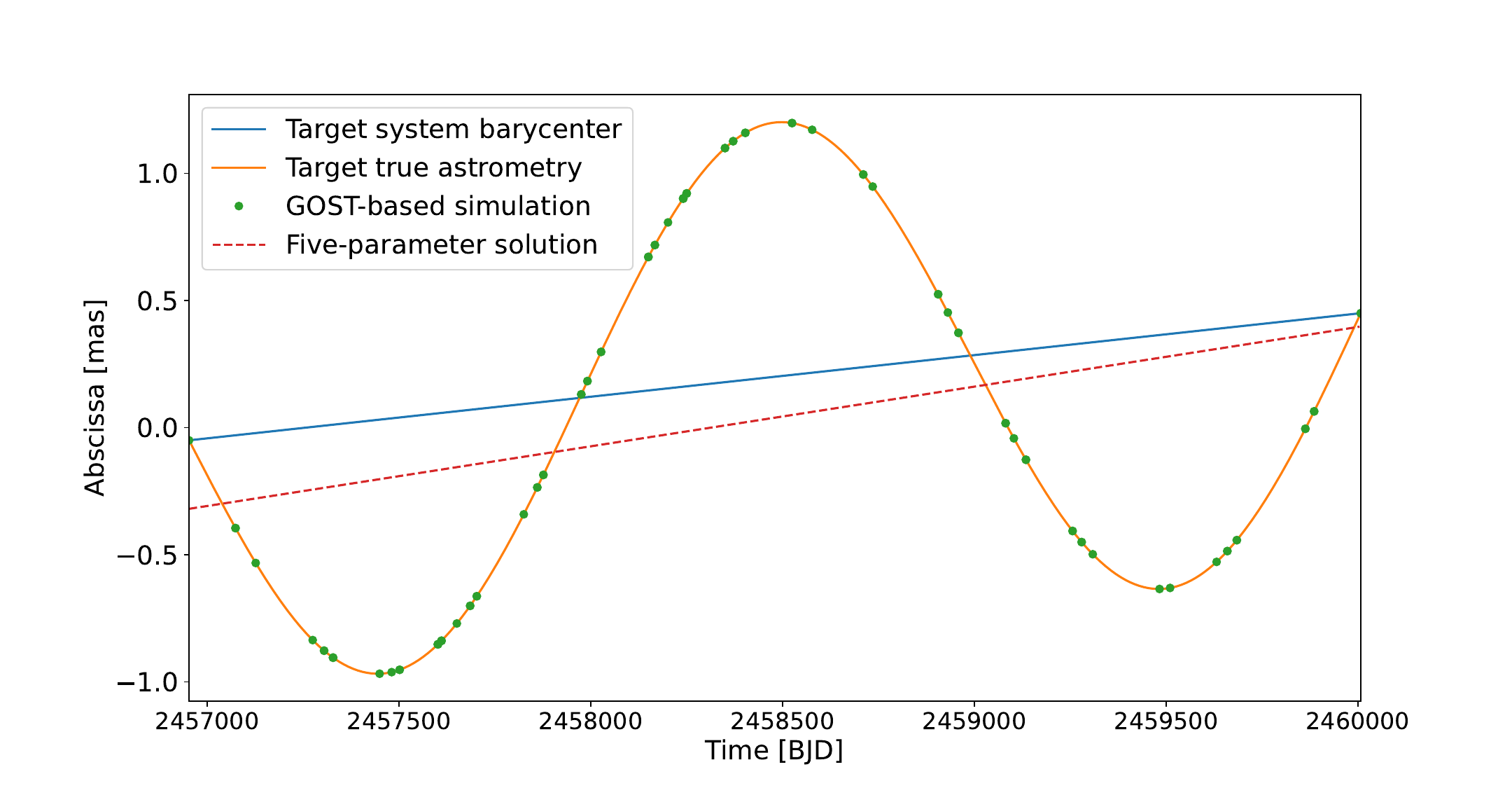}
        \begin{flushleft}
    \caption{Astrometric bias caused by utilization of a five-parameter astrometric model to fit the data of a target with an unidentified companion. The orange curve is simulated by accounting for both proper motion and reflex motion due to a long-period companion without considering parallax. The dashed red line represents the best-fit position and proper motion solution. The solid blue line represents the barycentric motion of this target.  The slope of a line is the proper-motion in this direction. The observation time comes from the Gaia GOST result of Barnard's star.}
    \label{fig:astrobias_gaia}
    \end{flushleft}
\end{figure*}

By assuming that the target is a single star, Gaia employs a five-parameter astrometric model to fit its raw data. The parameters of the model are right ascension (R.A.) denoted by $\alpha$, declination (Dec.) denoted by $\delta$, proper motion in R.A. denoted by $\mu_\alpha$, proper motion in Dec. denoted by $\mu_\delta$, and parallax denoted by $\varpi$.  This approach could lead to an astrometric bias in cases where an unknown companion is present, as shown in Fig. \ref{fig:astrobias_gaia}, or instrumental bias \citep{Lindegren_2021}. Only a small fraction of Gaia targets are analyzed without the single-star assumption in DR3 (443,205 in total 1.46 billion; \citealt{Gaia22b}). 

    Here is an order-of-magnitude estimation for this effect. The bias in the direction from the solar system barycenter to the target system barycenter (TSB) $\hat{n}$ having the bias $\Delta\hat{n}$ would lead to an RV ($v$) bias ($\Delta v$) of
\begin{equation}
    \Delta v\approx \Delta \hat{n}\cdot \vec{v}_{\oplus}\label{eq:oom}
\end{equation}
 where $\vec{v}_\oplus$ is Earth orbital velocity relative to the solar system barycenter as shown in Fig. \ref{fig:model}. This term would introduce an annual bias. %For a companion with mass $M_{\rm c}$ and semi-major axis $a_{\rm c}$, the 
%order-of-magnitude of the bias in the barycentric correction  $\Delta v_{max}=c\Delta z_B$ is

    %\begin{equation}
    %    \begin{aligned}
    %        \Delta v_{max,t}	\approx&2{\rm cm/s}\left(\frac{d}{1pc}\right)^{-1}\left(\frac{a_{\rm c}}{1AU}\right)^{-\frac{1}{2}}\left(\frac{M_{\rm c}}{M_J}\right)\left(\frac{M_s}{1M_\odot}\right)^{-\frac{1}{2}}\left(\frac{\Delta t}{20yr}\right) \\
    %        \Delta v_{max,r}\approx&2\mu {\rm m/s}\left(\frac{d}{1pc}\right)^{-1}\left(\frac{a_{\rm c}}{1AU}\right)^{-\frac{1}{2}}\left(\frac{M_{\rm c}}{M_J}\right)\left(\frac{M_s}{1M_\odot}\right)^{-\frac{1}{2}}\left(\frac{\Delta t}{20yr}\right)^2\left(\frac{|\mu|}{1000mas/yr}\right),
    %    \end{aligned}\label{eq:OOM}
    %\end{equation}
    %where the subscript $t$ of $\Delta v_{max}$ means the magnitude of annually periodic radial velocity noise caused by tangential bias (position or proper-motion) subscript $r$ stands for the RV bias caused by radial astrometric bias (parallax or systematic radial velocity); $d$ is the distance from the Sun to the target star; $M_s$ is the mass of target star, $|\mu|$ is the unbiased magnitude of the target star's proper-motion. In equation (\ref{eq:OOM}), the reference astrometry is assumed to be instantaneous epoch data, and the orbit of the target system is assumed to be edge-on. Since the Sun's radial velocity caused by the reflex motion of the Earth is no more than 10 cm/s, equation (\ref{eq:OOM}) shows that the correction for R. A., Dec. and proper-motion necessary for detecting Earth twins while the correction for parallax and systematic radial velocity is not necessary.

\begin{figure*}
        \centering
        \includegraphics[width=\linewidth]{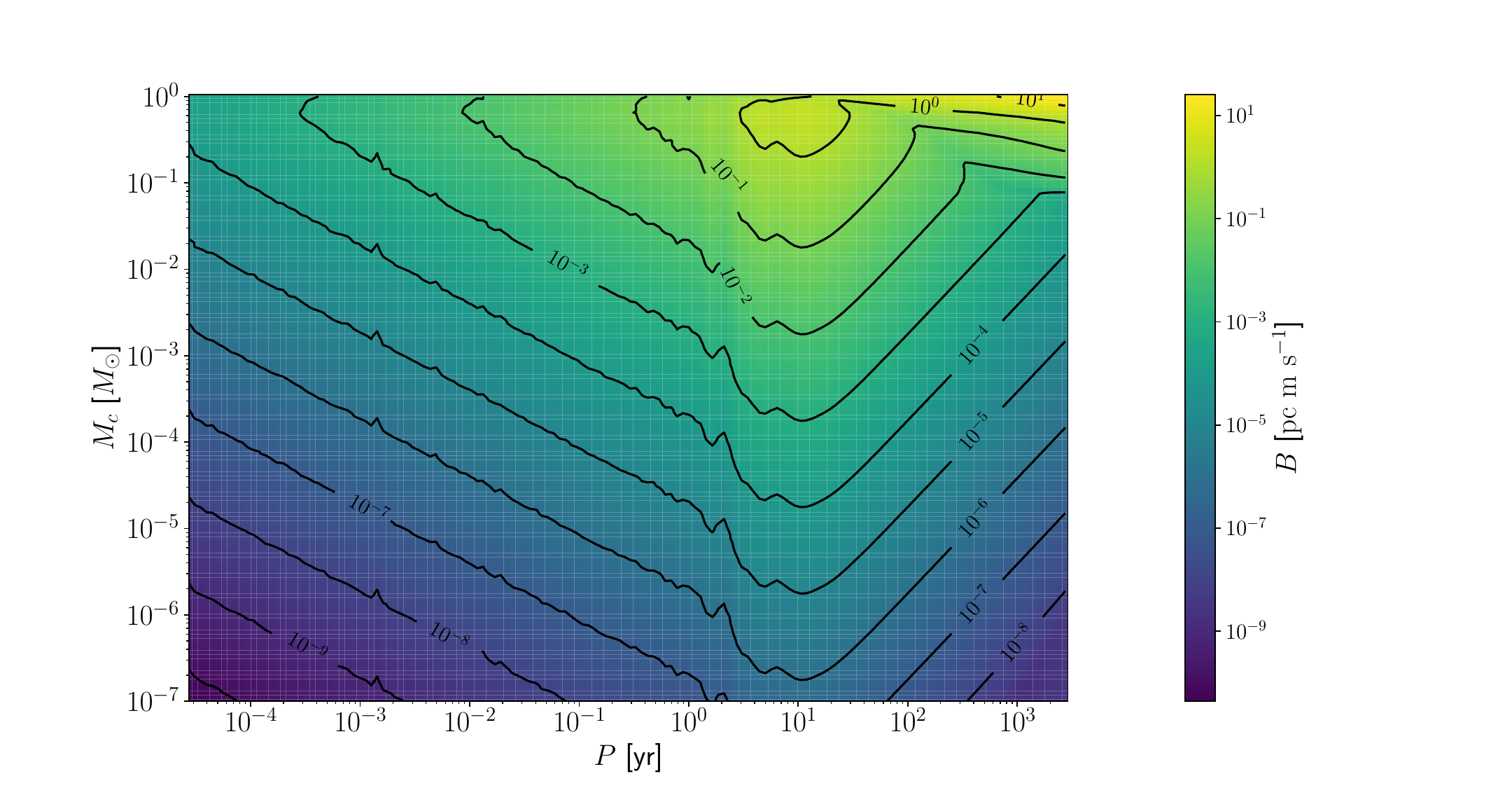}
        \caption{Magnitude of bias caused by an unresolved companion around a sun-like star. $P$ is the period of the massive companion; $M_c$ is the mass of the companion; $B$ is the scale-free maximum radial velocity bias normalized by distance. For a star at a distance of $d$, its radial velocity bias caused by the companion is $B/d$.%The red contour shows the RV bias of 0.1m/s at 10pc.The color for each $(M,P)$ represents the average radial velocity noise caused by the biased astrometry of the 1600 samples. 
        }
        \label{fig:m-p-Deltav}
    \end{figure*}

    To quantify the effect of astrometric bias, we conduct a simulation to assess the magnitude of it, as shown in Fig. \ref{fig:m-p-Deltav}. In this simulation, every target star is  assumed to have a mass of 1 $M_\odot$ and to
have an unresolved companion; The observational baseline of RV is assumed to be 10 years; The orbits of the systems are assumed to be circular; The target system's barycenter is assumed to have no proper-motion or parallax. The light emitted from the companions is considered in the simulation of photocenter motion. The G-band luminosity of the companions is calculated using the mass-luminosity relation given by \citealt{Pecaut13}. Based on Gaia sampling frequency, 36 points are sampled randomly in a 3 year observational baseline to simulate the along-scan position (abscissa) of Gaia source  \citep{GaiaDR3}. For a companion with a period $P$ and mass $M$, we uniformly sample its angular parameters and simulate its measured abscissae based on the photocenter of the system, and then fit a linear model to get the biased astrometry. The bias of relative radial velocity comes from the difference between the biased astrometry and the true astrometry of target star. Because the astrometric bias induced by companions is inversely proportional to the distance, the corresponding RV bias is also inversely proportional to the distance according to equation (\ref{eq:oom}). 
    %The results of Fig. \ref{fig:m-p-Deltav} in the vicinity of $P=5$yr match the prediction of equation (\ref{eq:OOM}). 
    
    It is shown in Fig. \ref{fig:m-p-Deltav} that this effect is most significant for target stars with long-period massive companions. This type of companions would lead to an annual RV bias of about 10\cms if the distance to the target system is 10\,pc. This effect is  suppressed for short-period  companions, because Gaia is more likely to  sample evenly within the phase of a short-period companion, leading to a fit with smaller bias. Meanwhile, this effect is slightly suppressed for stellar-mass companions. It is because the photocenter of the system would have a smaller nonlinear motion due to the light from the companion. However, for long-period stellar-mass companions, the light from the companion would make the system's photocenter far from the target star itself, leading to a significant bias.
    Because the astrometric bias is proportional to the parallax of the system, this effect is only significant for nearby stars.

%Considering that many targets are likely to host companions of the stellar or substellar kind, we mitigate the potential astrometric bias by considering astrometric offsets in the fitting of the model of barycentric motion to the measured RV data ($cz_{\rm meas}$). 

To model the impact of biased astrometry on the reduced RV, we calculate the Jacobian of RV relative to astrometry offset parameters. Through numerical calculation using PEXO, we obtain the Jacobian of the relative radial velocity of a given target with respect to the astrometric correction, which would be useful for correcting the input astrometry. The elements of the Jacobian ($\hat{J}$) can be written as
   \begin{equation}
       \hat{J}_{iA}\approx \frac{v_{i}(A_0+\Delta A)-v_{i}(A_0)}{\Delta  A},\label{eq:jacobian}
   \end{equation}
    where $v_{i}(A)$ is the radial velocity at time $t_i$ given astrometry $A$ where $i$ goes from 1 to the number of observation of this target $m$. $A$ denotes an astrometric parameter, which could be R. A., Dec., proper-motion in R. A. or proper-motion in Dec.. %Equation (\ref{eq:OOM}) indicates that the influence of parallax $\varpi$ and systematic radial velocity on barycentric correction results is not significant. 
    
    The impacts of potential bias in parallax ($\varpi$) and systematic radial velocity on barycentric correction is 
not considerred in this work. This is because the %radial astromeric bias (about $\sim$AU) is small relative to the distance to the target system (about $\sim$pc). It is smaller than the uncertainty of catalog parallax; The 
influence of parallax would not increase linearly with respect to time. The difference of secular acceleration caused by parallax bias is negligible for Earth-like planet detection. %Therefore, there is no need to correct them due to the Bayes factor concern, see section \ref{sec:correction}. Emperically, $\Delta A$ for R. A. and Dec. are chosen to be $0.01$mas and $\Delta A$ for proper-motion are chosen to be $0.001$mas/year.
    %It is the  numerical  realization for the first-order effects of biased astrometry.
 We assume that the scale of astrometric correction is small so that the change of radial velocity of every observation can be linearly approximated by the first order Taylor expansion. Our model of $\hat{\vec{v}}$ can be written as

   \begin{equation}
        \hat{\vec{v}}=\hat{J} \vec{a}+b\vec{1},
   \end{equation}
 where $\vec{a}=(\Delta \alpha, \Delta \delta,\Delta \mu_\alpha,\Delta \mu_\delta)^{\text{T}}$, is the amount of correction for this target; $b$ is the RV offset; $\vec{1}$ is a $m\cross 1$ vector which all entries equals to $1$.
   
\subsection{Correction of astrometric bias in a Bayesian framework}\label{sec:correction}

%The astrometry can be  optimized by using the weighted standard deviation of  relative radial velocity. 

%By assuming that different sources of noise are independent, we can decompose the weighted standard deviation of RV $\sigma_{v_{r }}^2$ into 

%\begin{equation}
%	\sigma_{v_{r }}^2=\sigma_{v_{ \text{star}}}^2+\sigma_{v_{ \text{barycentric}}}^2+\sigma_{v_{ \text{other source}}}^2,\label{eq:objopt}
%\end{equation} 
%explain different term
%where $\sigma_{v_{r }}$ is the weighted standard deviation of the observed RV data; $\sigma_{v_{ \text{star}}}$ is the weighted standard deviation due to the physics process of target stars; $\sigma_{v_{ \text{barycentric}}}$ is the error introduced by the barycentric correction. Equation (\ref{eq:objopt}) indicates that a more accurate barycentric correction would lead to a smaller standard deviation of the radial velocity data. We use this property to build our statistical model.

%\end{comment}

%The method which is similar to  Hipparcos astrometry correction \citep{1998HIPP} is used in this work. 
The likelihood of our model for one target star can be written as

\begin{equation}
	p(\vec{v}|\vec{a},b)=p(\vec{v}|\vec{\theta})=\frac{1}{(2\pi)^{\frac{m}{2}}|\Sigma|^{\frac{1}{2}}}e^{-\frac{1}{2}(\vec{v}-(\hat{J} \vec{a}+b\vec{1}))^\text{T}\Sigma^{-1}(\vec{v}-(\hat{J} \vec{a}+b\vec{1}))},\label{eq:model_crit}
\end{equation}
 where $\vec{v}$ is a $m \cross 1$ column vector, which contains the radial velocity data of this target star; $\vec{\theta}=(\vec{a}^{\rm T},b)^{\rm T}$ is a parameter set to be estimated in this model;  $\Sigma=\text{diag}(\sigma^2_1,\sigma^2_2,...,\sigma^2_m)$, where $\sigma_i$ is uncertainty of the $i$th radial velocity observation provided by HARPS; $|\Sigma|$ is the determinant of $\Sigma$.     The prior of our model can be written as 	
	\begin{equation}
		p(\vec{a},b)=p(\vec{\theta})=\frac{1}{(2\pi)^{\frac{m}{2}}|S|^{\frac{1}{2}}}e^{-\frac{1}{2}\vec{a}^\text{T}S^{-1}\vec{a}},
	\end{equation}
	where $S=\text{diag}(\sigma_\alpha^2,\sigma_\delta^2,\sigma_{\mu_\alpha}^2,\sigma_{\mu_\delta}^2)$ is the error bar given by Gaia DR3. According to Bayes' theorem, the posterior can be written as 	
	\begin{equation}
p(\vec{\theta}|\vec{v})=\frac{p(\vec{\theta})p(\vec{v}|\vec{\theta})}{p(\vec{v})}\propto p(\vec{\theta})p(\vec{v}|\vec{\theta}).
	\end{equation}
	To estimate $\vec{\theta}$, we employ the Maximum a posteriori (MAP) approach. By calculating the derivative of the logarithm of the posterior and equating it to zero, we obtain the solution 
		\begin{equation}
		\begin{pmatrix}
			\vec{a}\\b
		\end{pmatrix}=\begin{pmatrix}
			\hat{J}^\text{T}\Sigma^{-1}\hat{J} +S^{-1}& 	\hat{J}^\text{T}\Sigma^{-1}\vec{1}\\
			\vec{1}^{T}\Sigma^{-1}\hat{J} & \vec{1}^\text{T}\Sigma^{-1}\vec{1}
		\end{pmatrix}^{-1}\begin{pmatrix}
			\hat{J}^\text{T}\Sigma^{-1}\vec{v}\\
			\vec{v}^\text{T}\Sigma^{-1}\vec{1}
		\end{pmatrix}.\label{eq:res}
	\end{equation}
	 By marginalizing the posterior and using the property of multivariate normal distribution, we can obtain the covariance matrix of correction $\vec{a}$ by
	\begin{equation}
		\Sigma_{\vec{a}}=(\hat{J}^\text{T}\Sigma^{-1}\hat{J} +S^{-1})^{-1}\label{eq:covres}.
	\end{equation}
    The square root of the diagnal term of $\Sigma_{\vec{a}}$ is the uncertainty for each correction. By using equation (\ref{eq:res}) and (\ref{eq:covres}), we can obtain the correction $\vec{a}$ and its covariance $\Sigma_{\vec{a}}$. 
    
    Significance of the correction is measured by  Bayes factor (BF) estimated using Bayesian information criteria (BIC). BIC is used for Bayesian model selection. It is defined as 
    \begin{equation}
        {\rm BIC}=-2\ln{\hat{L}}+k\ln{m},\label{eq:defBIC}
    \end{equation} 
    where $\hat{L}$ is the maximized log likelihood; $k$ is number of parameters of the model. For the astrometric correction model which includes $\alpha$, $\delta$, $\mu_\alpha$, $\mu_\delta$, and $b$, we have $k=5$; For the model with no astrometric correction which only includes a constant offset $b$, we have $k=1$. Under the assumption of Laplace approximation \citep{Kass1995}, BF is derived from BIC as follows
    \begin{equation}
        \ln{\rm BF_{ij}}\approx \frac{\rm{BIC}_j-\rm{BIC}_i}{2},\label{eq:deflnBF}
    \end{equation}
    where $\rm BIC_i$ and $\rm BIC_j$ represents the BIC value of the compared model and the baseline model. In this research, we need to compare the following models: model 0 ({GDR2BarCor}), model 1 (GDR3PEXO) and model 2 (OptAstroPEXO).
      {GDR2BarCor} is the constant offset model based on HARPS RVbank data;  GDR3PEXO is the constant offset model based on PEXO reduced RV data which use Gaia DR3 astrometry;  OptAstroPEXO is the astrometric correction model which is described previously in this section{, see Table \ref{tab:models}}. We need to notice that  {GDR2BarCor} and  GDR3PEXO do not optimize astrometric parameters but only use different barycentric correction code and input astrometry, and  OptAstroPEXO is fitted to the RV data for optimizing astrometric parameters thus having 4 more parameters (change of R.A., Dec.{, proper motion offsets in R.A. and Dec.}) than  {GDR2BarCor} and GDR3PEXO. %{\color{red} In this work, the model using Gaia DR2 astrometry and PEXO reduced RV is not considered. This is because the bias induced by Barycorr is $\sim 0.01$\ms \citep{Wright14}, which is more than the typical magnitude of RV bias introduced by  the difference of Gaia DR3 and Gaia DR2 astrometric accuracy, see appendix \ref{app:pg}.}
      {There are variety of intermediate models between GDR3PEXO and GDR2BarCor which consider different barycentric correction code and input astrometry. They are not included in the model comparison framework due to their minimal differences with the GDR3PEXO model, as will be discussed in section \ref{subsec:pplinastro}.}
    
    The ``preferred model'' is defined as follows: 
\begin{itemize}
    \item  OptAstroPEXO is favored if $\rm lnBF_{20}>5$ and $\rm lnBF_{21}>5$.
    \item GDR3PEXO is favored if  $\rm lnBF_{10}>5$ and   $\rm lnBF_{21} < -5$.
    \item  {GDR2BarCor} is favored if  $\rm lnBF_{10} < -5$ and   $\rm lnBF_{20} < -5$.
    \item No model is favored if none the above rules are satisfied.
\end{itemize}
 These criteria are used because the condition 
lnBF>5 is a robust criterion for model comparison \citep{Kass1995}, and also see appendix \ref{app:pg}. Using these criteria, most targets don't have a favored model, see section \ref{sec:results}.

\begin{figure*}
        \centering
        \includegraphics[width=0.8\textwidth]{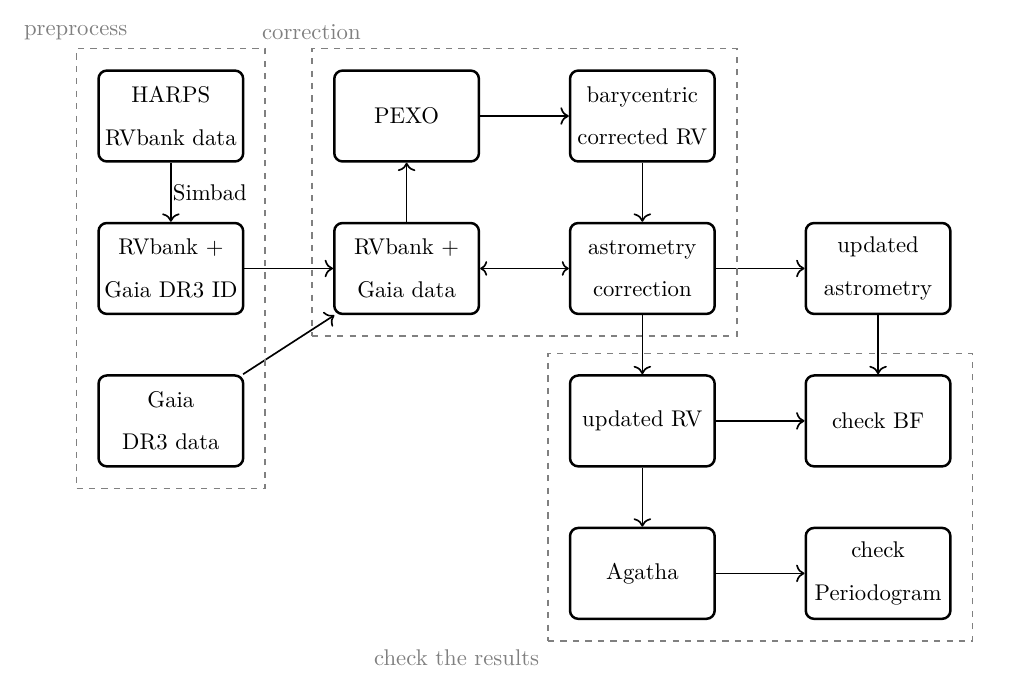}
        \caption{Workflow of the correction of the astrometric bias in RV data. }
        \label{fig:workflow} 
    \end{figure*}

    To evaluate the reduction in Earth motion after the astrometric correction, Bayes factor periodogram (BFP) is used to detect periodic signals in RV data. %There are two kinds of Bayesian periodogram: the Marginalized Likelihood Periodogram (MLP) and the Bayes Factor Periodogram (BFP) \citep{Agatha}. MLP is used for estimating the posterior of the signal frequency while BFP is used for signal detection. BFP is used in this work because we want to show that our method suppressed the residual of Earth motion in RV. 
    The value of the BF of periodogram is defined by equation (\ref{eq:defBIC}) and (\ref{eq:deflnBF}), where $\rm{BIC}_j$ and $\rm{BIC}_i$ represent the BIC of a pure noise model and a periodic model respectively. The criterion $\ln{\rm BF}=5$ is often chosen to be the threshold for significant periodic signals \citep{BFP5}.
     The BFP is calculated using Agatha \citep{Agatha}, which is a framework of periodograms with different noise models. Circular orbital model (the eccentricity is assumed to be 0) with white noise baseline model is used in this work. The workflow of the paper is shown in Fig. \ref{fig:workflow}.
    
\section{Results}\label{sec:results}
{In this section, we first compare intermediate models for the 266 HAPRS targets to investigate the impact of pipeline difference and input astrometry on the reduced RV data. We then compare GDR2BarCor (model 0), GDR3PEXO (model 1), and OptAstroPEXO (model 2) to study whether the optimization of input astrometry would improve the RV data reduction precision. Finally, we discuss the results for eight targets that show significantly improved RV precision after optimizing input astrometry or changing reduction pipeline.   }

    \subsection{{Model comparison}\label{subsec:generalres}\label{subsec:pplinastro}}
    
\begin{table}
{
        \caption{{Models used in this work}\label{tab:models}}
        \centering
        \begin{tabular}{lllll}
        \hline
         Model name&Input astrometry& Barycentric correction code   &Other name\\\hline
        GDR2BarCor & Gaia DR2$^{\rm (a)}$ & revised BarCor & Model 0\\
        GDR2Barycorr & Gaia DR2 & Barycorr$^{\rm (b)}$ &  --\\
        GDR2PEXO & Gaia DR2 & PEXO & --\\
        GDR3PEXO & Gaia DR3 & PEXO & Model 1\\
        OptAstroPEXO & Optimized astrometry & PEXO & Model 2\\\hline
        \end{tabular}
        
        \begin{flushleft}
            (a)  The input astrometry for most targets is from Gaia DR2, with several bright targets utilizing HIPPARCOS data. This practice is consistent across other models as well.\\
            (b) In this model, barycentric correction for the targets is re-conducted using the online Barycorr interface \url{https://github.com/tronsgaard/barycorr} \citep{Wright14}.
        \end{flushleft}}
\end{table}

\begin{figure}
    \centering
    \includegraphics[width = \linewidth]{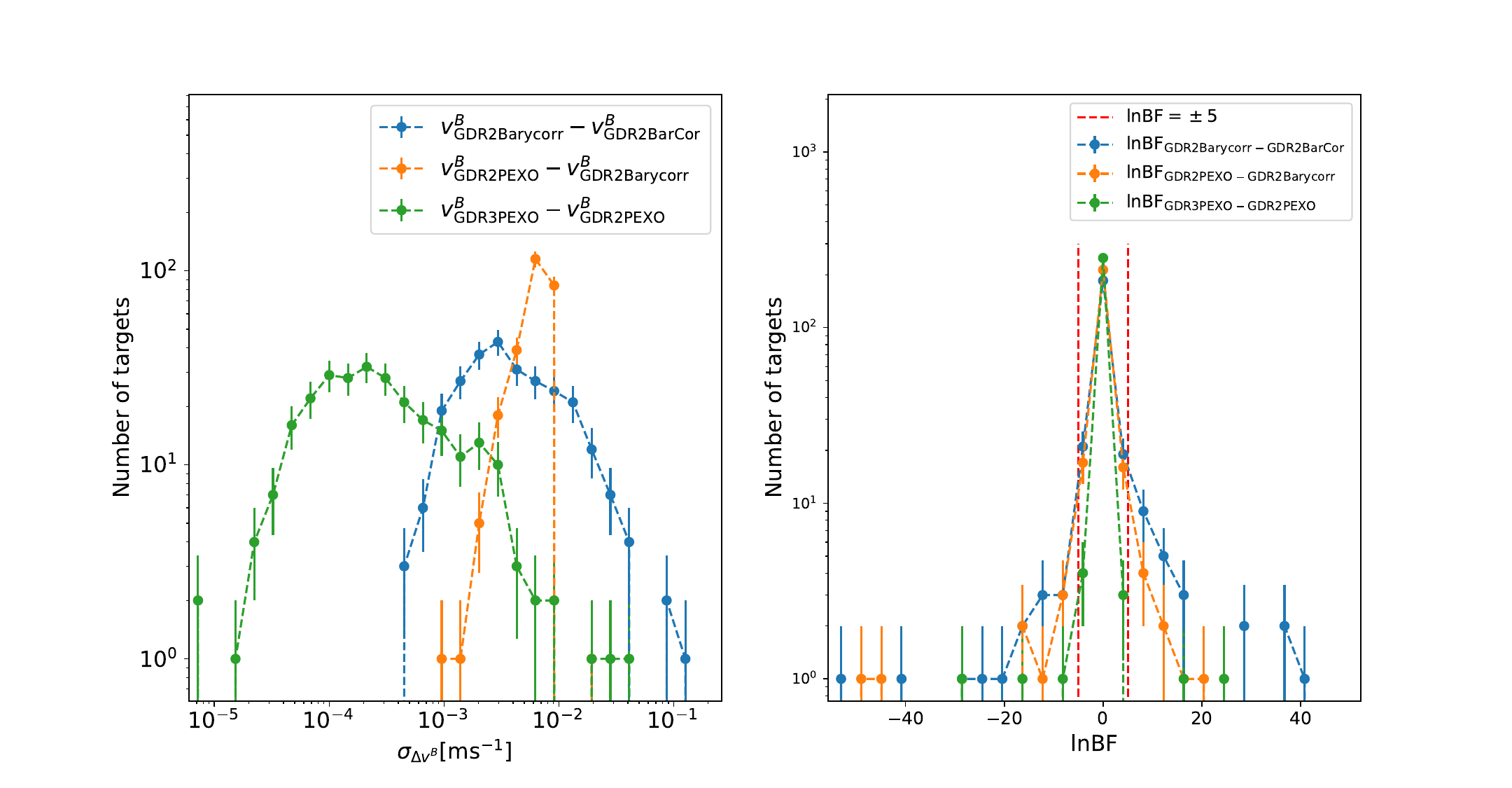}
    \caption{Comparison of intermediate models. 
\textbf{Left:} Distribution of the barycentric correction difference, where $v^B = cz_B$ represents the barycentric velocity obtained using the provided input astrometry and data reduction pipelines. $\sigma_{\Delta v^B}$ corresponds to the scattering of the difference in $v^B$
  for a target star. The error bars indicate uncertainties caused by poisson fluctuation. 
\textbf{Right:} Distribution of lnBFs when utilizing different input astrometry and data reduction pipelines. The red dashed lines are the thresholds for significant difference.}
    \label{fig:dzb}
\end{figure}

{
 We calculate the lnBFs of intermediate models and describe the model comparison shown in Fig. \ref{fig:dzb} as follows:
 \begin{itemize}
 \item{\bf GDR2BarCor and GDR2Barycorr.} These two models share the same input astrometry but with different barycentric correction codes. As seen in the right panel of Fig. \ref{fig:dzb}, the lnBF distribution (blue lines) is quite asymmetric, with 31 targets favoring GDR2Barycorr and 17 favoring GDR2BarCor. Such asymmetry is more than 3$\sigma$ away from Poisson noise and the preference for Barycorr over BarCor is likely caused by the fact that the BarCor does not account relativistic effects as Barycorr does. 
 The RV difference between these two models exceed 0.1\ms for 4 targets and exceed 0.01\ms for 55 targets, indicating the importance of pipeline selection in detecting Earth twins.
 \item{\bf GDR2Barycorr and GDR2PEXO.} The RV difference between these models are less than 1\cms, consistent with the 1\cms RV precision of GDR2Barycorr. Such difference does not lead to asymmetry in the lnBF distribution (orange line) shown in the right panel of Fig. \ref{fig:dzb}. Hence both PEXO and Barycorr are reliable in terms of conduct barycentric correction at the precision level of 1\cms. 
 \item {\bf GDR2PEXO and GDR3PEXO.} Only three targets shows RV difference of more than 1\cms due to change of Gaia DRs. The distribution of lnBF is symmetric, indicating little difference in barycentric correction due to the choice of Gaia DRs. 
 \end{itemize}}

	\begin{figure*}
		\centering
			\includegraphics[width=\textwidth]{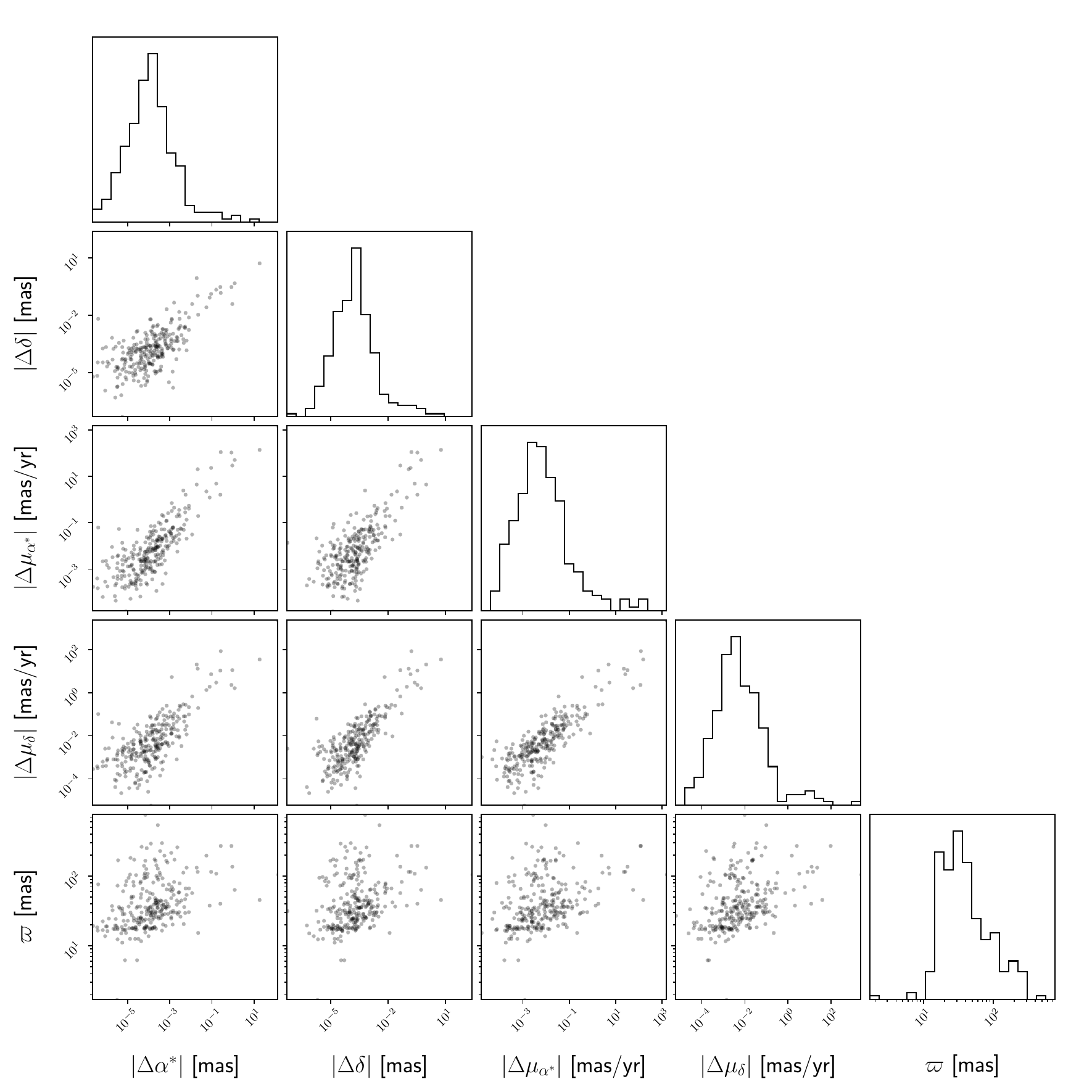}
			%\centering
			%\subcaption{correction for RA and DEC}
		\caption{Correlation between the astrometric correction and the parallax of the targets. Each point in the figure correspondence to a HARPS RVbank target. }
		\label{fig:astrocorr}

	\end{figure*}
	
	{After the comparison of intermediate models, we move on to investigate whether the optimized astrometry would improve the RV data reduction precision. Specifically, we compare OptAstroPEXO with GDR2BarCor and GDR3PEXO in the Bayesian framework to assess the impact of optimized astrometry. The distribution of inferred parameters of OptAstroPEXO combined with the parallax of the targets are shown in Fig. \ref{fig:astrocorr}.} It is apparent that the nearer targets have larger astrometric corrections. Meanwhile,  there is a positive correlation between the magnitude of correction for positions and proper motions. %$1642$ targets with radial velocity scatter less than 30$\rm m/s$ are displayed in this figure. This is because our method works for the targets without long-period companions. It is discussed in detail in section \ref{subsec:discussion}.
	
%	\begin{figure}
%	    \centering
%	    \includegraphics[width=0.95\textwidth]{RUWE_delta_ad}
%	    \includegraphics[width=0.95\textwidth]{RUWE_delta_pm}
%	    \caption{The relationship between the astrometry correction and the Gaia DR3 RUWE for targets with RV scatter less than 30m/s. The blue dots represent the stars which correction is larger than the standard deviation; the green line represents the result for the linear regression in log space. The Pearson correlation coefficient between the correction and RUWE is 0.2074 for the position and 0.5883 for the proper-motion.}
%	    \label{fig:RUWE_scale}
%	\end{figure}
 
	%To check the necessity of our correction with Gaia, we calculated the correlation between the correction and Unit Weight Error (RUWE), which measures the potential nonlinearity of stellar motion. %RUWE is proportional to the square root of the Gaia astrometry model's chi square. Large RUWE indicates that the astrometry of the target is not well modeled by Gaia's linear model, which may caused by the reflex motion with respect to its companion. Because our method is to correct the astrometry deviation caused by reflex motion, we expect that the scale of correction would be larger for the targets with larger REWE.
 %We show the relationship between Gaia DR3 RUWE and the offset parameters in Fig. \ref{fig:RUWE_scale}. We conduct a linear regression for the stars with RUWE greater than 1.4, which is considered to be a bad Gaia astrometric solution. The RUWE shows a significant positive correlation with the proper-motion correction.
	
	\begin{figure}
		\centering
            \includegraphics[width=\columnwidth]{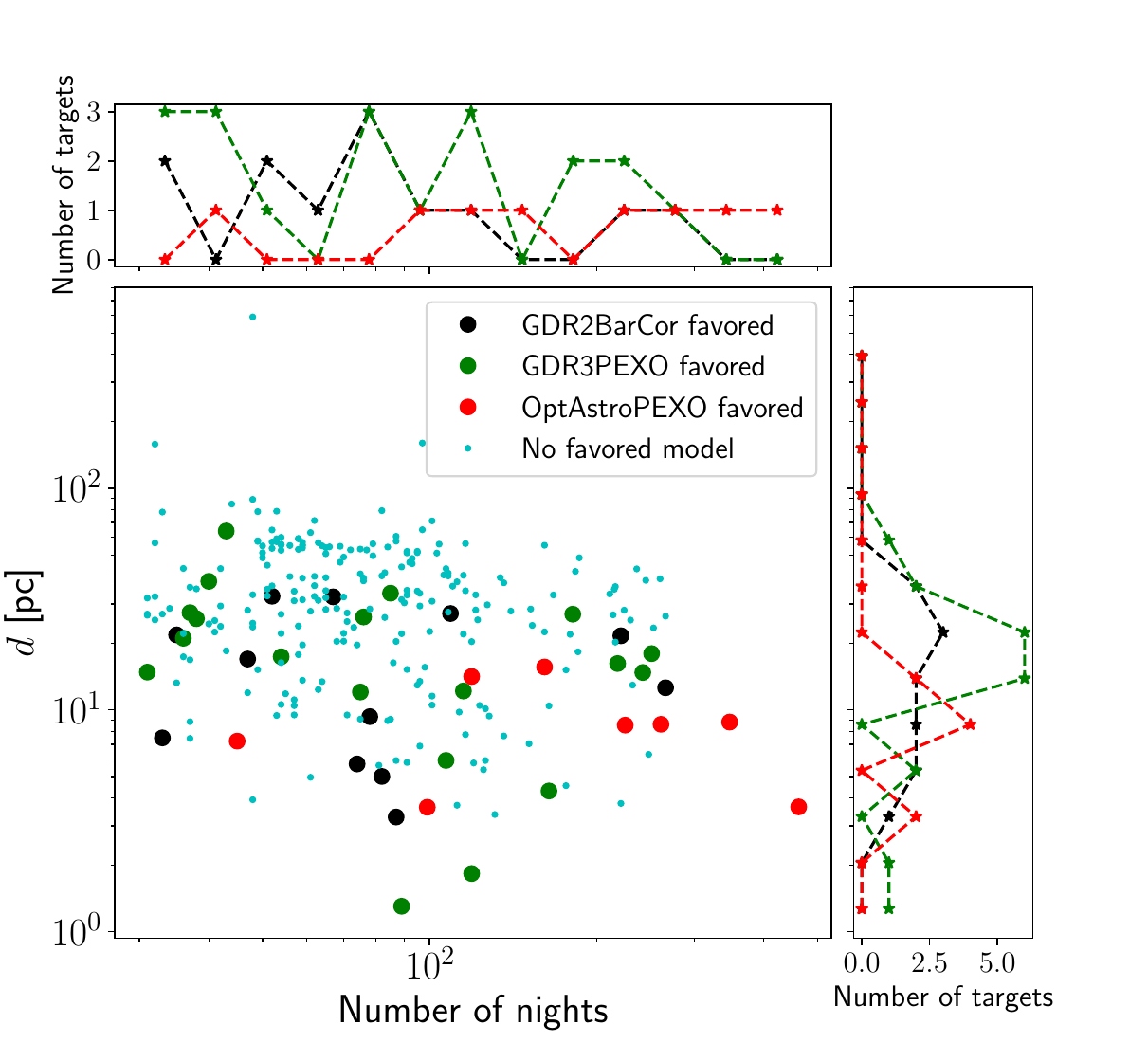}
		\caption{Distribution of selected targets on distance and number of nights with RV measurement. %\textbf{Upper panel}: 
  $d$ is the distance of the target stars obtained using parallax. The color of the points and histograms represents the favored model of the target. %\textbf{Lower panel}: The dots represent the target stars with both $\rm lnBF_{10}>0$ and $\rm lnBF_{20}>0$; the dashed lines are the thresholds for model selection. 
  }
		\label{fig:BIC}

	\end{figure}

We calculate the lnBFs for each target and model.  The results of model comparison and the magnitude of correction are shown in Table \ref{tab:statall}. Among the 266 targets,  13  of them favor {GDR2BarCor};  19 of them favor GDR3PEXO; 8 of them favor  OptAstroPEXO, where the ``favored model'' is defined in section \ref{sec:correction}. 
Considering the significance of RV bias correction, we find 141 targets having bias with slope of linear trend  with signal-to-noise ratio (SNR) greater than 3 when using GDR3PEXO, and 143 targets have this property when using  OptAstroPEXO. Among them, 3 targets have slope of linear trend  ($g_{10}$) higher than 0.01\,m\,s$^{-1}$\,yr$^{-1}$ using  GDR3PEXO, and 9 targets have slope of linear trend ($g_{20}$) higher than 0.01\,m\,s$^{-1}$\,yr$^{-1}$ using OptAstroPEXO. Meanwhile, we find 259 targets having an annual bias with an SNR greater than 3 when using GDR3PEXO or OptAstroPEXO. Among them, 11 targets have annual variation semi-amplitude ($K_{10}^{\rm annual}$) larger than 0.05\ms when using GDR3PEXO, which is comparable with the RV signal of Earth-like planets, and  14 targets have annual variation semi-amplitude ($K_{20}^{\rm annual}$) larger than 0.05\ms when using  OptAstroPEXO. Among this 14 targets, 10 of them are within the distance of 10\,pc. %We can see that the amount of change of position is highly correlated with the change of BIC which matches with the picture we described in Chapter \ref{sec:intro}. 

Fig. \ref{fig:BIC} shows the distance and the number of nights with RV observation for selected targets. The 8 targets that have significant astrometric correction (favor OptAstroPEXO) are closer than 20\,pc, and most of their RV data are obtained in more than 100 nights. The targets without significant astrometric correction %({\color{red} {\color{red}GDR2BarCor} model} or {\color{red} GDR3PEXO model}) 
do not display such preference. %Meanwhile, all the model 2 preferred targets are nearby stars closer than 20\,pc. %, which match the results in Fig. \ref{fig:m-p-Deltav}. 

\begin{table*}
\centering
\caption{Model comparison and RV variation caused by astrometric bias\label{tab:statall}. $g$ is the slope of linear trend  and $K^{\rm annual}$ is the semi-amplitude of the annual variation. The method for calculating $g$ and $K^{\rm annual}$ is shown in the appendix.}
\begin{tabular}{llr}
    \hline
    Criteria& Description  &    Number of targets\\
    \hline
    $\rm lnBF_{10} < -5$ and $\rm lnBF_{20} < -5$& {GDR2BarCor} favored        &  13  \\
    $\rm lnBF_{10}>5$ and $\rm lnBF_{21}< -5$&GDR3PEXO favored      &  19 \\
    $\rm lnBF_{20}>5$ and $\rm lnBF_{21}> 5$& OptAstroPEXO favored     &    8 \\

    ${\rm SNR}(K_{10}^{\rm annual})>3$&Significant  annual semi-amplitude of correction using  GDR3PEXO  &259\\
    ${\rm SNR}(K_{20}^{\rm annual})>3$&Significant  annual semi-amplitude of correction using  OptAstroPEXO &259\\
    ${\rm SNR}(K_{10}^{\rm annual})>3$ and $K_{10}^{\rm annual}$>0.05\ms & Annual semi-amplitude of correction >0.05\ms using  GDR3PEXO   &11 \\
    ${\rm SNR}(K_{20}^{\rm annual})>3$ and $K_{20}^{\rm annual}$>0.05\ms & Annual semi-amplitude of correction >0.05\ms using  OptAstroPEXO &14 \\
    ${\rm SNR}(g_{10})>3$&Significant  linear trend slope of correction using GDR3PEXO&141\\
    ${\rm SNR}(g_{20})>3$&Significant  linear trend slope of correction using OptAstroPEXO  &143\\
    ${\rm SNR}(g_{10})>3$ and $g_{10}$>0.01\,m\,s$^{-1}$\,yr$^{-1}$ &    Linear trend slope of correction >0.01\,m\,s$^{-1}$\,yr$^{-1}$ using  GDR3PEXO&3 \\
    ${\rm SNR}(g_{20})>3$ and $g_{20}$>0.01\,m\,s$^{-1}$\,yr$^{-1}$ &    Linear trend slope of correction >0.01\,m\,s$^{-1}$\,yr$^{-1}$ using OptAstroPEXO&9 \\

    \hline
\end{tabular}
\end{table*}

\begin{figure*}
	    \centering
	    \includegraphics[width=1\textwidth]{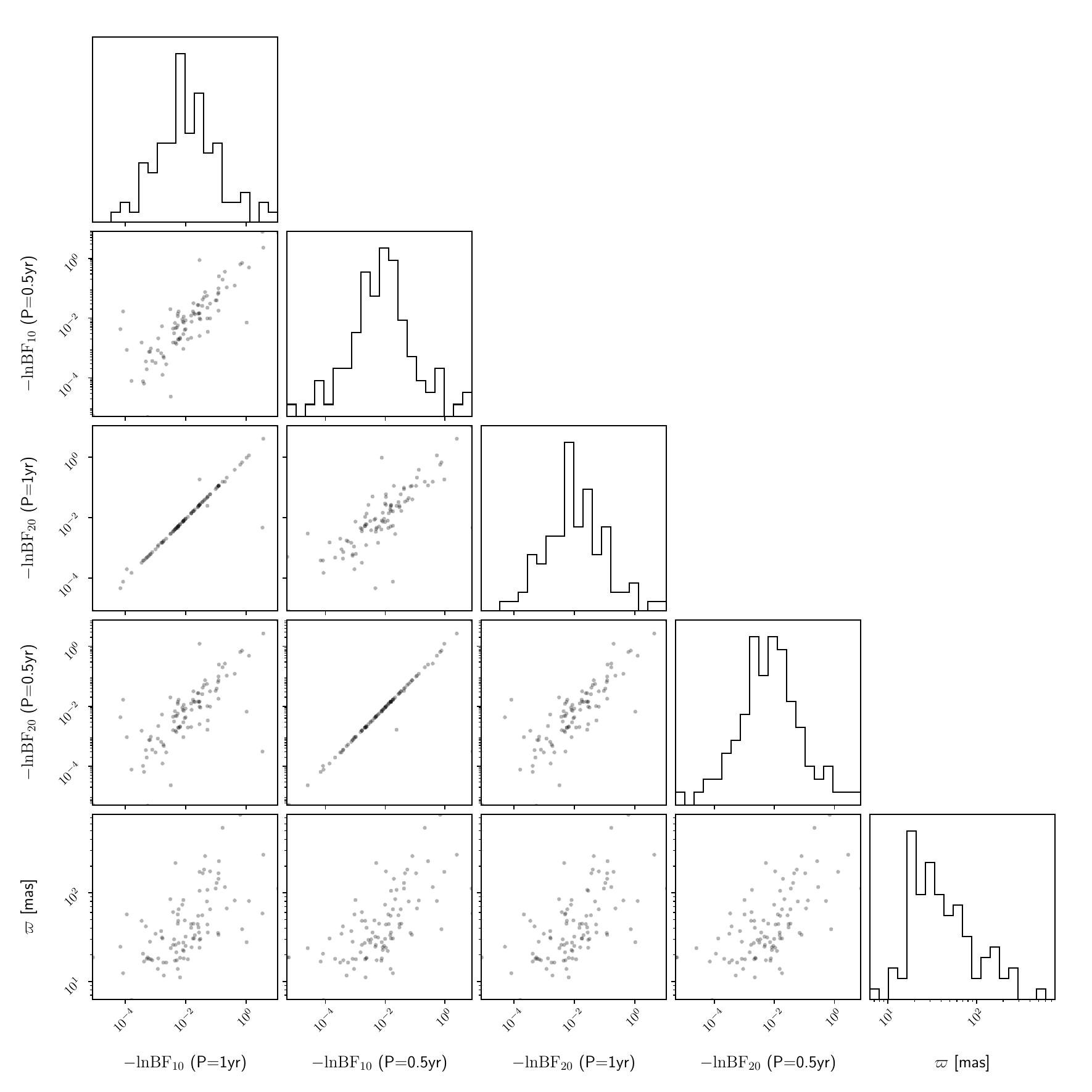}
	    \caption{Dependence of the BF of annual RV variation on parallax. $\varpi$ is the parallax of these targets from Gaia DR3.}
	    \label{fig:prga}
	\end{figure*}

	By calculating the BFPs of different models, we find that the annual signals of 96 targets are attenuated through the utilization of GDR3PEXO or OptAstroPEXO. The criterion for attenuation is defined as a decrease in the BF of the periodogram for signals corresponding to periods of 365.25 days and 182.625 days.  The decrease of BF of periodogram for these targets is shown in Fig. \ref{fig:prga}. It is shown that the decrease of the periodic signal is larger for nearby targets, which agrees with the intuition that astrometric bias is more significant for nearby targets. 
	
	%\begin{table}[]
	%    \centering
	%    \begin{tabular}{l|l}
	   %  Decrease of lnBFP at $P$=1yr   &  Number of target\\\hline
	   %      $0.01$&$312$\\
	     %    $0.1$&$198$\\
	     %    $1$&$122$\\
	     %    $10$&$47$
	  %  \end{tabular}
	%    \caption{Number of targets with decreased Bayes factor at 365.25 day after bias correction. }
	 %   \label{tab:dec_cor}
%	\end{table}

%	The decrease of BFP is shown in table \ref{tab:dec_cor}.
	
	According to equation (\ref{eq:oom}), the annual bias in barycentric correction is proportional to the line-of-sight direction. Because the bias of proper-motion would increase the bias of line-of-sight direction linearly with respect to the observational baseline $\Delta t$, it is expected that the correction for proper-motion should be larger than the correction for position, i.e. $\Delta t \sqrt{\Delta \mu_{\alpha^*}^2+\Delta \mu_\delta^2}>\sqrt{\Delta \alpha^{*2}+\Delta \delta^2}$. All the 266 targets satisfy this condition.%$2413$ targets in $2692$ targets satisfy $\Delta t \sqrt{\Delta \mu_{\alpha^*}^2+\Delta \mu_\delta^2}>\sqrt{\Delta \alpha^{*2}+\Delta \delta^2}$. It is shown in Fig. \ref{fig:cpradepm}. A more detailed discussion on the validity of the correction is presented in section \ref{subsec:discussion}.
	
	%\begin{figure}
	%    \centering
	%    \includegraphics[width=\linewidth]{cpradepm}
	%    \caption{Comparison between the positional correction and the proper-motional correction. $2413$ targets in $2692$ targets satisfy $\Delta t \sqrt{\Delta \mu_{\alpha^*}^2+\Delta \mu_\delta^2}>\sqrt{\Delta \alpha^{*2}+\Delta \delta^2}$.}
	%    \label{fig:cpradepm}
	%\end{figure}

    \subsection{Targets with significant astrometric correction}

 Among the 266 selected targets, there are 8 targets having significant astrometric bias. They are GJ 105 A, GJ 17 ($\zeta$ Tuc), GJ 506 (61 Vir), GJ 616 (18 Sco), GJ 691 ($\mu$ Ara), GJ 785, GJ 845 ($\epsilon$ Ind) and HD 10700 ($\tau$ Cet). Their astrometric correction together with their statistics are shown in Table \ref{tab:astrocorr}. We also show the results of GJ 699 (Barnard's star) for comparison. The default parameters used below are from Gaia DR3.

    \begin{itemize}
    \item  \textbf{GJ 105 A.} GJ 105 is a triple star system 7.2\,pc away from the Earth. % \citep{Feng_2021}. 
    GJ 105 A is a K-type star while GJ 105 B and GJ 105 C are M-dwarfs. GJ 105 A is expected to have a large correction because GJ 105 C is an astrometrically unresolved companion with a period for about $P=76$\,yr \citep{Feng_2021}. According to Fig. \ref{fig:m-p-Deltav}, the RV bias caused by GJ 105 C is $\sim 0.1$\ms. Meanwhile, the  semi-amplitude of the annual variation of this target is $K_{20}^{\rm annual}=6.28\pm 0.12$\cms, which is comparable to the signal of Earth-like planets and matches the simulation results. 

    \item \textbf{GJ 17.} GJ 17 is a F-type star 8.6\,pc away from the Earth. It is included in the target star list of the HWO \citep{Mamajek23}. No exoplanets or stellar companions are found around it.  However, a significant astrometric correction is shown for this target. The Gaia EDR3 proper-motion anomaly \citep{Kervella22} shows that it possibly has a companion with a mass larger than one Jupiter mass, which may account for this astrometric bias. The Gaia DR3 RUWE for this target is larger than 1.4, indicating the existance of unknown companions \citep{GaiaDR3}. The RV bias of it might thus come from the bias of Gaia astrometry. The  semi-amplitude of the annual variation of this target is $K_{20}^{\rm annual}=5.00\pm 0.02$\cms, which is comparable to the signal of Earth-like planets.
    
    \item \textbf{61 Vir (GJ 506).} 61 Virginis (GJ 506) is a G-type star 8.5\,pc away from the Earth. It is included in the target star list of the HWO \citep{Mamajek23}. 61 Virginis has a confirmed super-Earth 61 Vir b whose orbital period is $P_{\rm b}=4.2150\pm0.0001$\,d, and mass is $M_{\rm b}\ge 5.98^{+0.3}_{-0.29}\,M_\oplus$. It also has two confirmed Neptune mass planets 61 Vir c and 61 Vir d whose periods are $P_{\rm c}=38.079\pm 0.008$\,d and $P_{\rm d}=123.2\pm 0.2$\,d, and mass are $M_{\rm c}\ge 17.94^{+0.73}_{-0.7}\,M_\oplus$ and $M_{\rm d}\ge 10.82^{+1.23}_{-1.03}\,M_\oplus$ respectively \citep{Katherine23}. However, these planets can not account for the significant RV bias. The G-band magnitude of it is 4.5\,mag. Such a bright target would lead to pixel saturation which can not be fully mitigated by reducing integration time controlled by blocking gates. This issue may lead to astrometric bias \citep{Lindegren_2020}. 
    
    The correction results for this target is shown in Fig. \ref{fig:GJ506corr}. The top left panel shows that the RV data is well sampled in the observational baseline of $\sim 10$\,yr; The bottom left panel shows that the pattern for RV correction is an annual signal superposed on a linear trend, which is a typical pattern for correction for all the targets; The top right panel shows that the signals of all its confirmed planets are significant; The bottom right panel shows that the annual signals are attenuated after the correction. Meanwhile, the significance of all confirmed exoplanets is enhanced after using the corrected astrometry. This is because the reduction in annual signals increase the relative significance of non-annual signals, including the signals of planets.

    \begin{figure*}
        \centering
        \includegraphics[width=\linewidth]{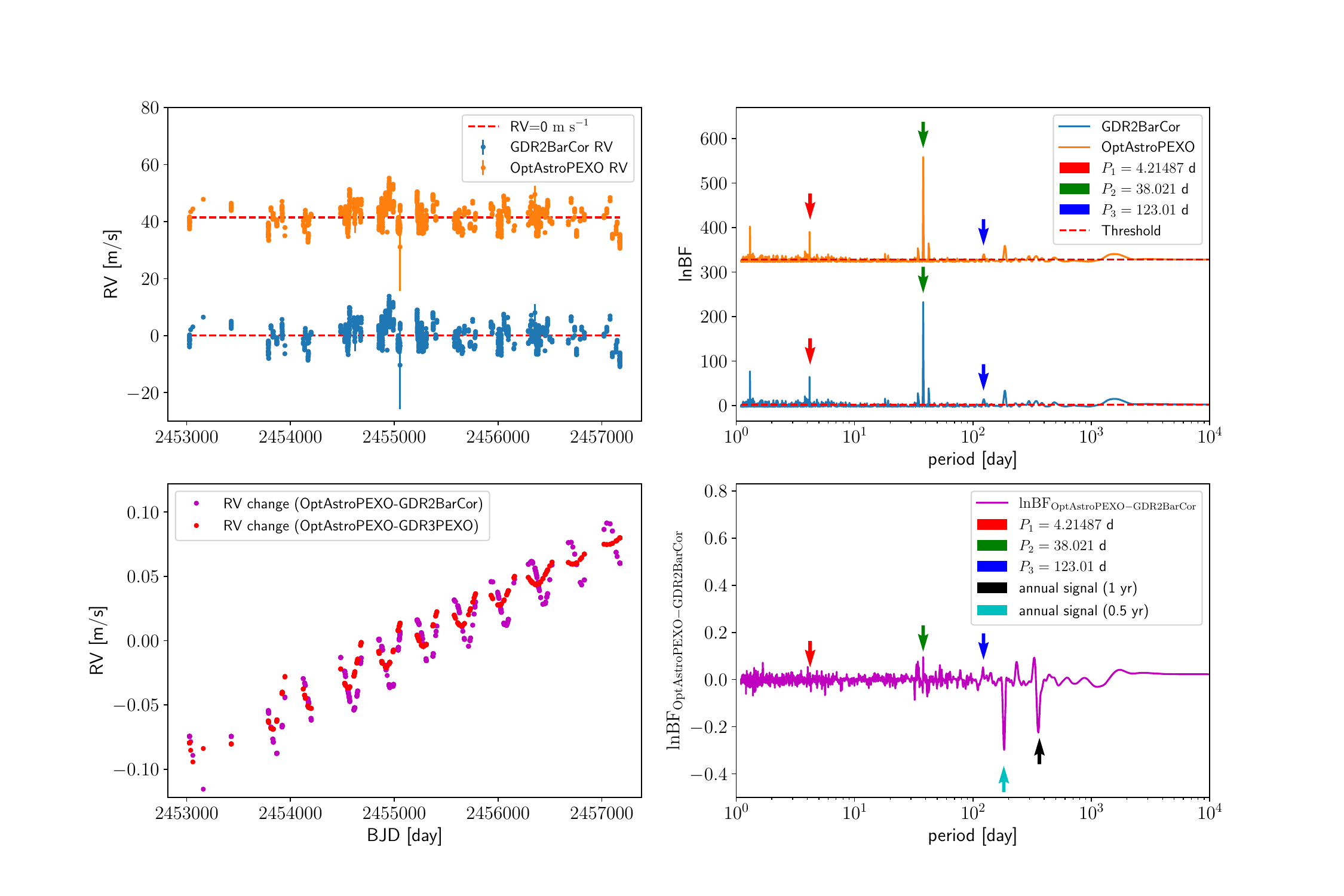}
        \caption{Radial velocity correction and BFP for 61 Virginis. \textbf{Top left:} The radial velocity data before and after the correction. An offset of 41.4\ms is applied to the corrected RV data. \textbf{Bottom left:} The change of radial velocity after the astrometric correction. The median is subtracted for comparision. \textbf{Top right:} The BFP of radial velocity before the correction and after the astrometric correction. The arrows represent the periods of confirmed planets. An offset of 325.6 is applied to the BFP of the corrected RV. \textbf{Bottom right:} The change of BFP after the correction. The median is subtracted for comparison. The significance of all three planets is increased while the significance of annual signals is decreased.}\label{fig:GJ506corr}
    \end{figure*}

    \item \textbf{GJ 616.} GJ 616 is a G-type star 14\,pc away from the Earth. It is included in the target star list of the HWO \citep{Mamajek23}.
    % \citep{Keenan89}. 
    \citealt{Katherine23} find a super Earth candidate using radial velocity method. The mass for this exoplanet candidate is $M\ge 6.77\pm0.86 M_\oplus$ and its period is $P=19.8777\pm0.0062$\,d. The G-band magnitude of it is 5.33\,mag, which is probably suffering from the systematics of bright stars. Meanwhile, there exists possibility that the astrometric bias is attributed to an unidentified long-period massive companion.

    \item \textbf{GJ 691.} GJ 691 ($\mu$ Arae) is a G-type star 15.6\,pc away from the Earth. It is included in the target star list of the HWO \citep{Mamajek23}.
    It has 4 confirmed exoplanets detected using radial velocity \citep{pepe07}. Its heaviest planet GJ 691 e is a planet with about 1.97 Jupiter mass and a period of 4019 days, which is too small to account for the astrometric bias and radial velocity mitigation. The G-band magnitude for this object is 4.94\,mag, which could potentially introduce astrometric bias. The  semi-amplitude of the annual variation of this target is $K_{20}^{\rm annual}=8.37\pm 0.06$\cms.

    \item \textbf{GJ 785.} GJ 785 is a K-type star 8.8\,pc away from the Earth. It is included in the target star list of the HWO \citep{Mamajek23}. %\citep{NSTARS}. 
    It has two confirmed Neptune-sized planets whose orbital periods are $P_{\rm b}=74.278\pm 0.035$\,d and $P_{\rm c}=549.1\pm 4.5$\,d, and  mass is $M_{\rm b}\ge 14.28^{+0.64}_{-0.63}\, M_\oplus$ and $M_{\rm c}\ge 14.96^{+1.21}_{-1.18}\, M_\oplus$ respectively \citep{pepe11}. These companions can not explain the astrometric bias. The G-band magnitude of it is 5.5\,mag, which indicates that the bias may due to the Gaia astrometry. 

   \item \textbf{GJ 845.} GJ 845 is a K-type star 3.6\,pc away from the Earth. It is included in the target star list of the HWO and it is considered to be a good (tier A) target \citep{Mamajek23}.
   %\citep{NSTARS}. 
   It have a binary brown dwarf companion and a planet of about 3 Jupiter mass \citep{Feng19p}. According to Fig. \ref{fig:m-p-Deltav}, these companions would lead to an RV bias of $\sim 0.1$\ms, which is comparable with the annual signals of Earth-like planets. Meanwhile, the  semi-amplitude of the annual variation of this target is $K_{20}^{\rm annual}=15.89\pm 0.4$\cms, which is consistent with the results of Fig. \ref{fig:m-p-Deltav}. This indicates that the bias of GJ 845 is caused by its companions. %Meanwhile, the annual RV signal is less significant after the astrometric correction.%The astrometric correction give the astrometry of the system barycenter, which is important for long-baseline radial velocity modeling.

    \item \textbf{$\tau$ Ceti (HD 10700).} $\tau$ Ceti (HD 10700) is a G-type star 3.7\,pc away from the Earth %\citep{Keenan89} 
    with 4  planets \citep{Feng_2017}.  It is included in the target star list of the HWO \citep{Mamajek23}. % Its G-band magnitude is 3.3\,mag \citep{GaiaDR3}, which  is so bright that may lead to a bias in Gaia astrometry. 
    No known planets could induce such bias.
    This target have RUWE=2.6343, which is larger than 1.4, indicating strong systematics of the Gaia astrometry. Meanwhile, the  semi-amplitude of the annual variation of this target is $K_{20}^{\rm annual}=10.77\pm 0.15$\cms, which is comparable with the annual signals of Earth-like planets.
    
    \item \textbf{Barnard's star (GJ 699).} Barnard's star is an M-dwarf 1.8\,pc away from the Earth. %\citep{Gizis_1997}. 
    \citealt{GJ699b} claims the existence of a super-Earth planet orbiting near the
    snow line of Barnard’s star which period is $P=233$\,d. %The parameters of the proposed Barnard's star b in shown in table (\ref{tab:barnardplanet}).
     The input astrometry of Gaia DR3 % is better than DR2 , and 
      does not have any significant bias. %The change of input astrometry from Gaia DR2 to Gaia DR3 for this target is shown in Table \ref{tab:astrocorr}.
     %The difference between Gaia DR2 and Gaia DR3 for Barnard's star is about $1.3\, \rm mas\, yr^{-1}$ for proper-motion in R.A., which can explain 
%We need to notice that there seems to be a huge difference in the position between Gaia DR2 and Gaia DR3. It is because the reference epoch of Gaia DR3 is half year later than Gaia DR2 \citep{GaiaDR3}. This difference is due to the proper motion of Barnard's star. 
     The corrected RV and the BFP is shown in Fig. \ref{fig:GJ699corr}.  The RV difference is mainly caused by the utilization of PEXO. It is shown that the periodic signal of the exoplanet candidate is more significant while the annual alias due to the Earth motion is reduced when using PEXO. 
%    \begin{comment}
%        \begin{table}
%        \centering
%        \begin{tabular}{l|l|l|l|l}
%             Planet name& Minimun mass $m\sin i$($M_\oplus$)& Orbital period $P$(days)&Semi-major axis $a$(AU)& %Eccentricity $e$ \\\hline
%             Barnard's star b& $3.23\pm 0.44$&$232.80^{+0.38}_
%{ -−0.41}$&$0.404\pm 0.018$&$0.32^{+0.10}_
%{ -−0.15}$
%        \end{tabular}
%        \caption{Exoplanet candidate of Barnard's star}
%        \label{tab:barnardplanet}
%    \end{table} 
%    \end{comment}

    \begin{figure*}
        \centering
        \includegraphics[width=\linewidth]{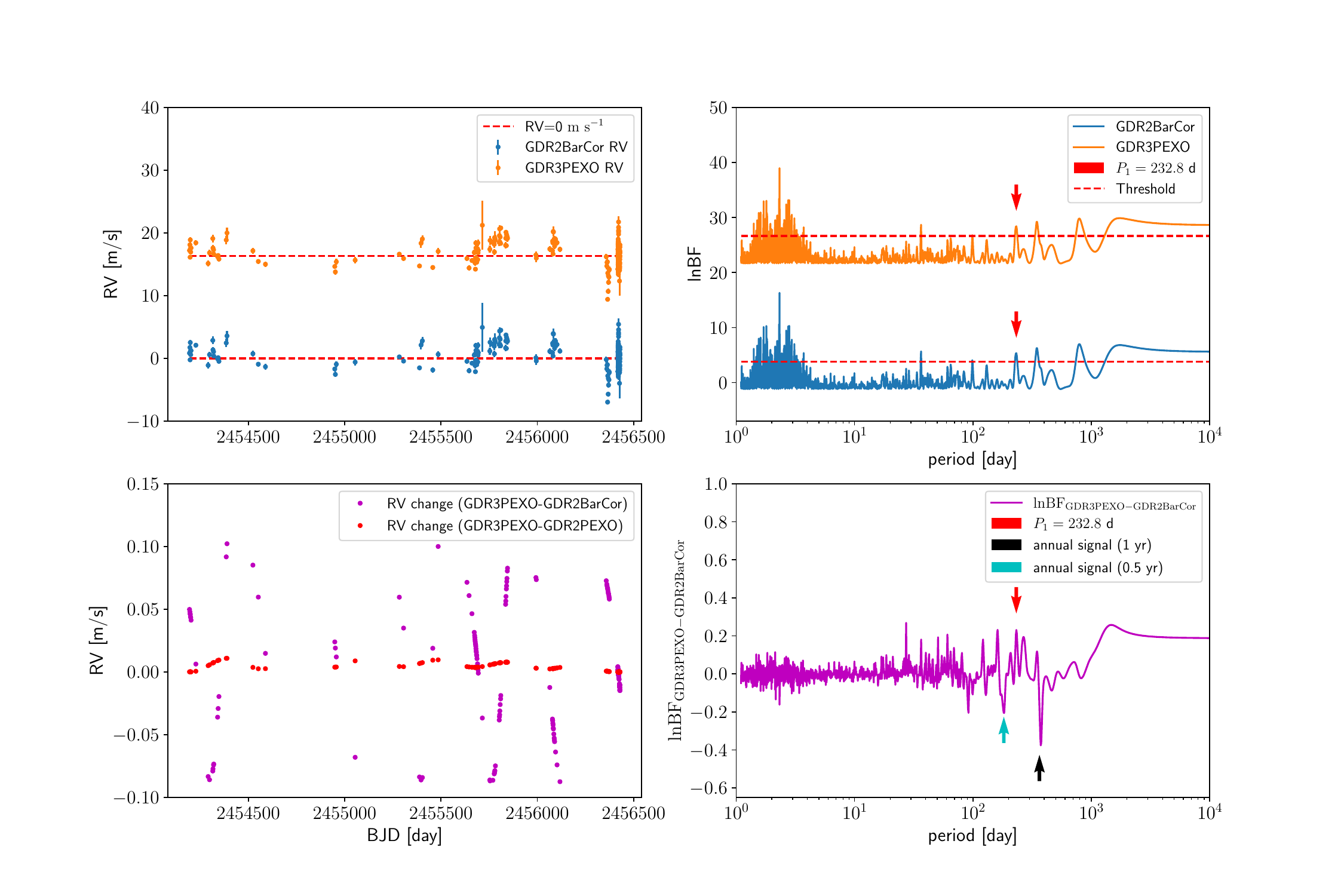}
        \caption{Change of radial velocity and the BFP when using GDR3PEXO. Configurations of this figure is similar to Fig. \ref{fig:GJ506corr}, but for Barnard’s star. In the top left panel, an offset of 16.3\ms is applied for the corrected RV. In the top right panel, an offset of 22.8 is applied for the BFP after correction.}
        \label{fig:GJ699corr}
    \end{figure*}
    \end{itemize}

\begin{landscape}
\begin{table}
        \caption{Inferred quantities for selected targets. 
        %For model 2 preferred targets, their astrometric correction is relative to Gaia DR3; For the model 1 preferred targets, their astrometric correction is the change of Gaia DR3 relative to Gaia DR2.  
        $i$ is the preferred model; $g$ is the linear trend slope of the change of radial velocity corrected by the preferred model, i.e. the slope of the bottom left plot of Fig. 
        \ref{fig:GJ506corr}; $K^{\rm annual}$ represents the annual variation semi-amplitude of the preferred model; $\rm lnBF_{i0}$ is the lnBF comparing model i  with the baseline model using RVbank. The results for the remaining targets are accessible to the public. A comprehensive description of the data product is shown in the appendix. }
   \centering
 
    \begin{tabular}{lrrrrrrrrrr}
\hline
 Target & Model $i$ & $\Delta \alpha^*$       & $\Delta \delta$      & $\Delta \mu_{\alpha^*}$ & $\Delta\mu_\delta$ &  $g_{i0}$ &  $K^{\rm annual}_{i0}$ &   $\rm lnBF_{i0}$&RUWE\\
  &&mas&mas&mas\,yr$^{-1}$&mas\,yr$^{-1}$&$\rm cm\,s^{-1}\,yr^{-1}$&$\rm cm\,s^{-1} $&&\\\hline
  GJ 105 A &      OptAstroPEXO &    $-0.87\pm 3.19$&   $-0.04\pm 2.17$ &    $-30.2\pm 3.3$ &    $-12.0\pm 2.3$  &        $ 2.11\pm 0.04$  & $ 6.28\pm 0.12$   &     12.0  &  7.3\\
   GJ 17 &      OptAstroPEXO &    $-0.02\pm 0.48$ &    $0.11\pm 1.17$  &     $21.7\pm 1.4$&     $13.6\pm 1.5$  &        $-2.06\pm 0.00$  & $ 5.00\pm 0.02$   &     241.2 &  3.2\\
  GJ 506 &      OptAstroPEXO &    $-0.09\pm 0.98 $&    $0.14\pm 1.10$ &     $24.1693\pm 1.3$ &      $7.8\pm 1.2$  &        $ 1.58\pm 0.01$  & $ 2.90\pm 0.05$   &     427.0 &  1.2\\
  GJ 616 &      OptAstroPEXO &    $-0.00\pm 0.65$ &   $-0.01\pm 0.55$ &     $-0.3\pm 0.7$ &      $5.6\pm 0.7$  &        $ 0.14\pm 0.00$  & $ 0.19\pm 0.03$   &     43.4  &  1.0\\
  GJ 691 &      OptAstroPEXO &    $-1.16\pm 0.38$ &   $-0.48\pm  0.55$ &     $53.9\pm 0.8$&      $1.7\pm 0.6$  &        $ 0.40\pm 0.02$  & $ 8.37\pm 0.06$   &    4390.7 &  0.9\\
  GJ 785 &      OptAstroPEXO &     $0.00\pm 0.39$ &   $-0.00\pm 0.30$ &      $2.5\pm 0.5$ &     $-0.3\pm 0.3$  &        $-0.17\pm 0.01$  & $ 2.09\pm 0.01$   &     79.5  &  1.0\\
  GJ 845 &      OptAstroPEXO &    $-0.25\pm 0.38$ &   $-0.15\pm 0.66$ &    $118.5\pm 0.8$ &    $-90.6\pm 0.9$  &        $-10.83\pm 0.05$  & $ 15.89\pm 0.40$   &   30901.3 &  1.1\\
HD 10700 &      OptAstroPEXO &     $0.81\pm 1.14$ &    $0.31\pm 0.83$ &   $-110.7\pm 0.9$ &      $2.4\pm 0.9$  &        $-4.72\pm 0.00$  & $ 10.77\pm 0.15$   &    5357.2 &  2.6\\
Barnard's star &      GDR3PEXO &     --- &          --- &             ---&                  --- &        $ 0.15\pm 0.06$  & $ 10.93\pm 0.57$   &     13.8  &  1.1\\
\hline
\end{tabular}

        \label{tab:astrocorr}
\end{table}
\end{landscape}

\section{Discussion and conclusion}\label{sec:cad}

We introduce a PEXO-based method to mitigate astrometric bias in barycentric correction for the detection of Earth-like planets. By applying this method to 266 HARPS targets, we find that  8 targets, including GJ 105 A, GJ 17, GJ 506, GJ 616, GJ 691, GJ 785, GJ 845 and HD 10700, have significantly biased astrometry. All of them are within a  distance of 20\,pc. The RV bias in GJ 105 A and GJ 845, characterized by a strong linear trend and an annual bias of  $\sim 0.1$\ms, is likely attributed to their known companions. For GJ 17, GJ 506, GJ 616, GJ 691, GJ 785 and HD 10700, they are bright targets without known massive companions. Their astrometric bias could be attributed either to the Gaia instrumental systematics or to unidentified long-period massive companions. We show the details of radial velocity correction and periodogram analysis for GJ 506.  The significance of all its confirmed exoplanets is enhanced after the astrometric correction, because the astrometric correction reduce the RV systematis as well as the annual variation.

{For the remaining 226 targets, neither input nor optimized astrometry is preferred probably due to lack of well-sampled high precision RVs over long time span.} 
The periodogram analyses reveal that the annual RV variation of 96 targets is decreased when PEXO in combination with GDR3 or corrected astrometry. 14 among the 266 targets have annual bias of at least 0.05\ms using corrected astrometry, and 10 of them are within the distance of 10\,pc. Our successful correction of these bias demonstrate the importance of unbiased barycentric correction in detecting Earth-like planets.

{Based on a comparison of intermediate models that share the same input astrometry but have different barycentric correction code, the Barycorr and PEXO pipelines have at most 1\cms difference in barycentric correction and both are precise enough for detecting Earth-twin signals. Both of them are favored over the BarCor pipeline that is used in SERVAL by \cite{Trifonov20} to process RVbank data. Because BarCor does not account for relativistic effects, it can introduce sub-\ms RV bias in barycentric correction. In particular, we show that the barycentric correction for Barnard’s star can be significantly improved by replacing BarCor with PEXO. Therefore, we highly recommend using relativistic pipelines such as Barycorr and PEXO for barycentric correction when detecting Earth twins.}

 In future work, a simultaneous modeling of barycentric and reflex motion is needed to totally remove the assumptions of barycentric correction in the RV analyses, as implemented in tools such as TEMPO2 \citep{edwards06} and PEXO \citep{Feng19}. 

\section*{Acknowledgements}

This work is supported by Shanghai Jiao Tong University 2030 Initiative.  We thank the referee for valuable comments, which help improving the manuscript. 
This work has made use of data from the European Space Agency (ESA) mission Gaia (\url{https://www.cosmos.esa.int/gaia}), processed by the Gaia Data Processing and Analysis Consortium (DPAC, \url{https://www.cosmos.esa.int/web/gaia/dpac/consortium}). Funding for the DPAC has been provided by national institutions, in particular the institutions participating in the Gaia Multilateral Agreement. This research has made use of the NASA Exoplanet Archive (\url{https://exoplanetarchive.ipac.caltech.edu}), which is operated by the California Institute of Technology, under contract with the National Aeronautics and Space Administration under the Exoplanet Exploration Program. This research has made use of  the SIMBAD database (\url{http://simbad.u-strasbg.fr/simbad}), operated at CDS, Strasbourg, France. This work have used TOPCAT, a starlink table/VOTable processing software. We are deeply grateful to the HARPS team at Observatoire de
Genève, Observatoire de Haute-Provence, Laboratoire d’Astrophysique de Marseille, Service d’Aéronomie du CNRS, Physikalisches Institut de Universität
Bern, ESO La Silla, and ESO Garching, who built and maintained the HARPS instrument, and were generous enough to make the data public. This work is based on observations collected at the European Organization for Astronomical Research in the Southern
Hemisphere under ESO programmes: 0100.C-0097, 0100.C-0111, 0100.C-0414,
0100.C-0474, 0100.C-0487, 0100.C-0750, 0100.C-0808, 0100.C-0836, 0100.C0847, 0100.C-0884, 0100.C-0888, 0100.D-0444, 0100.D-0717, 0101.C-0232,
0101.C-0274, 0101.C-0275, 0101.C-0379, 0101.C-0407, 0101.C-0516, 0101.C0829, 0101.D-0717, 0102.C-0338, 0102.D-0717, 0103.C-0548, 0103.D-0717,
060.A-9036, 060.A-9700, 072.C-0096, 072.C-0388, 072.C-0488, 072.C-0513,
072.C-0636, 072.D-0286, 072.D-0419, 072.D-0707, 073.A-0041, 073.C-0733,
073.C-0784, 073.D-0038, 073.D-0136, 073.D-0527, 073.D-0578, 073.D-0590,
074.C-0012, 074.C-0037, 074.C-0102, 074.C-0364, 074.D-0131, 074.D-0380,
075.C-0140, 075.C-0202, 075.C-0234, 075.C-0332, 075.C-0689, 075.C-0710,
075.D-0194, 075.D-0600, 075.D-0614, 075.D-0760, 075.D-0800, 076.C-0010,
076.C-0073, 076.C-0155, 076.C-0279, 076.C-0429, 076.C-0878, 076.D-0103,
076.D-0130, 076.D-0158, 076.D-0207, 077.C-0012, 077.C-0080, 077.C-0101,
077.C-0295, 077.C-0364, 077.C-0530, 077.D-0085, 077.D-0498, 077.D-0633,
077.D-0720, 078.C-0037, 078.C-0044, 078.C-0133, 078.C-0209, 078.C-0233,
078.C-0403, 078.C-0751, 078.C-0833, 078.D-0067, 078.D-0071, 078.D-0245,
078.D-0299, 078.D-0492, 079.C-0046, 079.C-0127, 079.C-0170, 079.C-0329,
079.C-0463, 079.C-0488, 079.C-0657, 079.C-0681, 079.C-0828, 079.C-0927,
079.D-0009, 079.D-0075, 079.D-0118, 079.D-0160, 079.D-0462, 079.D-0466,
080.C-0032, 080.C-0071, 080.C-0664, 080.C-0712, 080.D-0047, 080.D-0086,
080.D-0151, 080.D-0318, 080.D-0347, 080.D-0408, 081.C-0034, 081.C-0119,
081.C-0148, 081.C-0211, 081.C-0388, 081.C-0774, 081.C-0779, 081.C-0802,
081.C-0842, 081.D-0008, 081.D-0065, 081.D-0109, 081.D-0531, 081.D-0610,
081.D-0870, 082.B-0610, 082.C-0040, 082.C-0212, 082.C-0308, 082.C-0312,
082.C-0315, 082.C-0333, 082.C-0357, 082.C-0390, 082.C-0412, 082.C-0427,
082.C-0608, 082.C-0718, 083.C-0186, 083.C-0413, 083.C-0794, 083.C-1001,083.D-0668, 084.C-0185, 084.C-0228, 084.C-0229, 084.C-1039, 085.C-0019,
085.C-0063, 085.C-0318, 085.C-0393, 086.C-0145, 086.C-0230, 086.C-0284,
086.D-0240, 087.C-0012, 087.C-0368, 087.C-0649, 087.C-0831, 087.C-0990,
087.D-0511, 088.C-0011, 088.C-0323, 088.C-0353, 088.C-0513, 088.C-0662,
089.C-0006, 089.C-0050, 089.C-0151, 089.C-0415, 089.C-0497, 089.C-0732,
089.C-0739, 090.C-0395, 090.C-0421, 090.C-0540, 090.C-0849, 091.C-0034,
091.C-0184, 091.C-0271, 091.C-0438, 091.C-0456, 091.C-0471, 091.C-0844,
091.C-0853, 091.C-0866, 091.C-0936, 091.D-0469, 092.C-0282, 092.C-0454,
092.C-0579, 092.C-0721, 092.C-0832, 092.D-0261, 093.C-0062, 093.C-0409,
093.C-0417, 093.C-0474, 093.C-0919, 094.C-0090, 094.C-0297, 094.C-0428,
094.C-0797, 094.C-0894, 094.C-0901, 094.C-0946, 094.D-0056, 094.D-0596,
095.C-0040, 095.C-0105, 095.C-0367, 095.C-0551, 095.C-0718, 095.C-0799,
095.C-0947, 095.D-0026, 095.D-0717, 096.C-0053, 096.C-0082, 096.C-0183,
096.C-0210, 096.C-0331, 096.C-0417, 096.C-0460, 096.C-0499, 096.C-0657,
096.C-0708, 096.C-0762, 096.C-0876, 096.D-0402, 096.D-0717, 097.C-0021,
097.C-0090, 097.C-0390, 097.C-0434, 097.C-0561, 097.C-0571, 097.C-0864,
097.C-0948, 097.C-1025, 097.D-0156, 097.D-0717, 098.C-0269, 098.C-0292,
098.C-0304, 098.C-0366, 098-C-0518, 098.C-0518, 098.C-0739, 098.C-0820,
098.C-0860, 098.D-0717, 099.C-0093, 099.C-0138, 099.C-0205, 099.C-0303,
099.C-0304, 099.C-0374, 099.C-0458, 099.C-0491, 099.C-0798, 099.C-0880,
099.C-0898, 099.D-0717, 1101.C-0721, 180.C-0886, 183.C-0437, 183.C-0972,
183.D-0729, 184.C-0639, 184.C-0815, 185.D-0056, 188.C-0265, 188.C-0779,
190.C-0027, 191.C-0505, 191.C-0873, 192.C-0224, 192.C-0852, 196.C-0042,
196.C-1006, 198.C-0169, 198.C-0836, 198.C-0838, 281.D-5052, 281.D-5053,
282.C-5034, 282.C-5036, 282.D-5006, 283.C-5017, 283.C-5022, 288.C-5010,
292.C-5004, 295.C-5031, 495.L-0963, 60.A-9036, 60.A-9700, and 63.A-9036.

%%%%%%%%%%%%%%%%%%%%%%%%%%%%%%%%%%%%%%%%%%%%%%%%%%
\section*{Data Availability}

The data underlying this article are available in the article and in its online supplementary material.



\bibliographystyle{mnras}
\bibliography{example.bib} % if your bibtex file is called example.bib

%\begin{thebibliography}{99}
%\end{thebibliography}

%%%%%%%%%%%%%%%%%%%%%%%%%%%%%%%%%%%%%%%%%%%%%%%%%%

%%%%%%%%%%%%%%%%% APPENDICES %%%%%%%%%%%%%%%%%%%%%

\appendix
\section{Headers for the result data set}
\subsection{Corrected astrometry table}

\begin{table*}
    \caption{Header for the new astrometry table}
    \label{tab:headerastrometry}
    \centering
    \begin{tabular}{lll}
    \hline
         Column name &Description &Unit \\
         \hline
         target&Name of the target &---\\
         RA&Corrected Gaia DR3 R. A. in ICRS frame&deg\\
         dRA*&Astrometric correction for R.A. times $\cos\delta$&deg\\
         DE&Corrected Gaia DR3 Dec. in ICRS frame &deg\\
         dDE&Astrometric correction for Dec. &deg\\
         pmRA&Corrected Gaia DR3 R. A. proper-motion&mas\,yr$^{-1}$\\
         dpmRA&Astrometric correction for proper motion in  R.A. direction &mas\,yr$^{-1}$\\
         pmDE&Corrected Gaia DR3 Dec. proper-motion&mas\,yr$^{-1}$\\
         dpmDE&Astrometric correction for proper motion in  Dec. direction &mas\,yr$^{-1}$\\
         source&Source of astrometric data (Gaia DR3 /DR2 /HIPPARCOS) &---\\
         std\_delta\_ra&Error bar for the R. A. corection&deg\\
         std\_delta\_de&Error bar for the Dec. corection&deg\\
         std\_delta\_pmra&Error bar for the R. A. proper-motion correction&mas\,yr$^{-1}$\\
         std\_delta\_pmde&Error bar for the DE proper-motion correction&mas\,yr$^{-1}$\\
         cov\_delta\_ra\_de&Covariance between R. A. correction and Dec. correction&$\rm deg^2$\\
         cov\_delta\_ra\_pmra&Covariance between R. A. correction and R. A. proper-motion correction&$\rm deg\,mas\,yr^{-1}$\\
         cov\_delta\_ra\_pmde& Covariance between R. A. correction and Dec. proper-motion correction&$\rm deg\,mas\,yr^{-1}$\\
         cov\_delta\_de\_pmra&Covariance between Dec. correction and R. A. proper-motion correction&$\rm deg\,mas\,yr^{-1}$\\
         cov\_delta\_de\_pmde&Covariance between Dec. correction and Dec. proper-motion correction&$\rm deg\,mas\,yr^{-1}$\\
         cov\_delta\_pmra\_pmde&Covariance between R. A. proper-motion correction and Dec. proper-motion correction&$\rm mas^2\,yr^{-2}$\\
         lnBFi0&lnBF between model i and  {GDR2BarCor} &---\\
         PMcorr\_larger&Whether $\Delta t \sqrt{\Delta \mu_{\alpha^*}^2+\Delta \mu_\delta^2}>\sqrt{\Delta \alpha^{*2}+\Delta \delta^2}$ &---\\
        dlnBF10\_annual&Change of lnBF for periodogram using GDR3PEXO at period equals to one year&---\\
         n\_Obs& RV epoch number  &---\\
         baseline&The observation baseline of HARPS radial velocity  &day\\
         scatter&The weighted standard deviation of RV &\ms\\
         coverage&Number of nights with RV observation&---\\
         pmRA\_HG& The R.A. proper-motion from the HGCA catalog &mas\,yr$^{-1}$\\
         pmDE\_HG& The Dec. proper-motion from the HGCA catalog &mas\,yr$^{-1}$\\
         e\_pmRA\_HG& The error of R.A. proper-motion from the HGCA catalog &mas\,yr$^{-1}$\\
         e\_pmDE\_HG& The error of Dec. proper-motion from the HGCA catalog &mas\,yr$^{-1}$\\
         g\_ij&The linear trend slope of RV difference between model $i$ and model $j$&$\rm m\,s^{-1}\,yr^{-1}$\\
         K\_i0&The annual variation semi-amplitude of residual of RV difference between model i and  {GDR2BarCor}  &\ms\\
         sigma\_mi&The standard deviation of reduced RV for model i &\ms\\
         excessive\_noise&Excessive noise from Gaia DR3&---\\
         RUWE&Gaia DR3 RUWE&---\\
\hline
    \end{tabular}
\end{table*}
\newpage
\subsection{Corrected radial velocity table}
Some entries of this table is the same as the HARPS RVbank data set. 
\begin{table*}
    \caption{Header for the new RV table}
    \label{tab:headerrv}
    \centering
    \begin{tabular}{lll}
        \hline
        Column name &Description &Unit \\\hline
          BJD(TDB)\_2 & The radial velocity observational epoch derived using PEXO and astrometric correction&day\\
         BERV\_2&The PEXO barycentric correction results for the RV after the astromerty correction&\ms\\
         BJD(TDB)\_1 & The radial velocity observational epoch derived using PEXO and Gaia DR3 astrometry&day\\
         BERV\_1 &The PEXO barycentric correction results for the RV before the astromerty correction&\ms\\ 
         RV\_2&The PEXO relative RV results after the astromerty correction&\ms\\
         RV\_1&The PEXO relative RV results using Gaia DR3 astrometry&\ms\\
         \hline
    \end{tabular}
\end{table*}
\section{Linear trend slope and the annual variation semi-amplitude of the RV bias}

In section \ref{subsec:generalres}, we show the linear trend slope $g$ and the annual variation semi-amplitude $K^{\rm annual}$ of the RV bias. In this appendix, we show the method for calculating these quantities and estimating their uncertainties.

Because the linear trend is coupled with the annual signal, it is better to consider these effects simultaneously. The correction of the radial velocity $\Delta v_r(t)$ can be modeled using a linear combination of the annual signal and a linear trend, which is
\begin{equation}
    \Delta \hat{v}_r=A\cos(2\pi \frac{t}{1\rm yr})+B\sin(2\pi \frac{t}{1\rm yr})+C t +D,
\end{equation}
where the annual variation semi-amplitude $K^{\rm annual}=\sqrt{A^2+B^2}$; the linear trend slope $g=C$. The least square solution for $\hat{\theta}=(A,B,C,D)^T$ is
\begin{equation}
    X=\begin{pmatrix}
        \cos(2\pi \frac{t_1}{1\rm yr})&\sin(2\pi \frac{t_1}{1\rm yr})&t_1&1\\
        \cos(2\pi \frac{t_2}{1\rm yr})&\sin(2\pi \frac{t_2}{1\rm yr})&t_2&1\\
        \vdots&\vdots&\vdots&\vdots\\
        \cos(2\pi \frac{t_m}{1\rm yr})&\sin(2\pi \frac{t_m}{1\rm yr})&t_m&1
    \end{pmatrix},\quad \Delta\vec{v}=\begin{pmatrix}
        \Delta v_{1}\\
        \Delta v_{2}\\
        \vdots\\
        \Delta v_{m}
    \end{pmatrix}, \quad \hat{\theta}=(X^{\rm T} X)^{-1}X^{\rm T} \Delta \vec{v}
\end{equation}
where $t_i$ is the time for the $i$th RV data; $\Delta v_{i}$ is the bias of RV data at $t_i$.

To estimate the uncertainties of $g$ and $K$, we conduct a boostrap with 500 resamples. For each sample, we repeat the previous illustrated process to obtain their $g$ and $K^{\rm annual}$. The uncertainty of  $g$ and $K^{\rm annual}$ is estimated using the standard deviation of the ensemble.

\section{The distribution of $\rm\ln BF_{10}$  }\label{app:pg}

\begin{figure}
		\centering
            \includegraphics[width=\columnwidth]{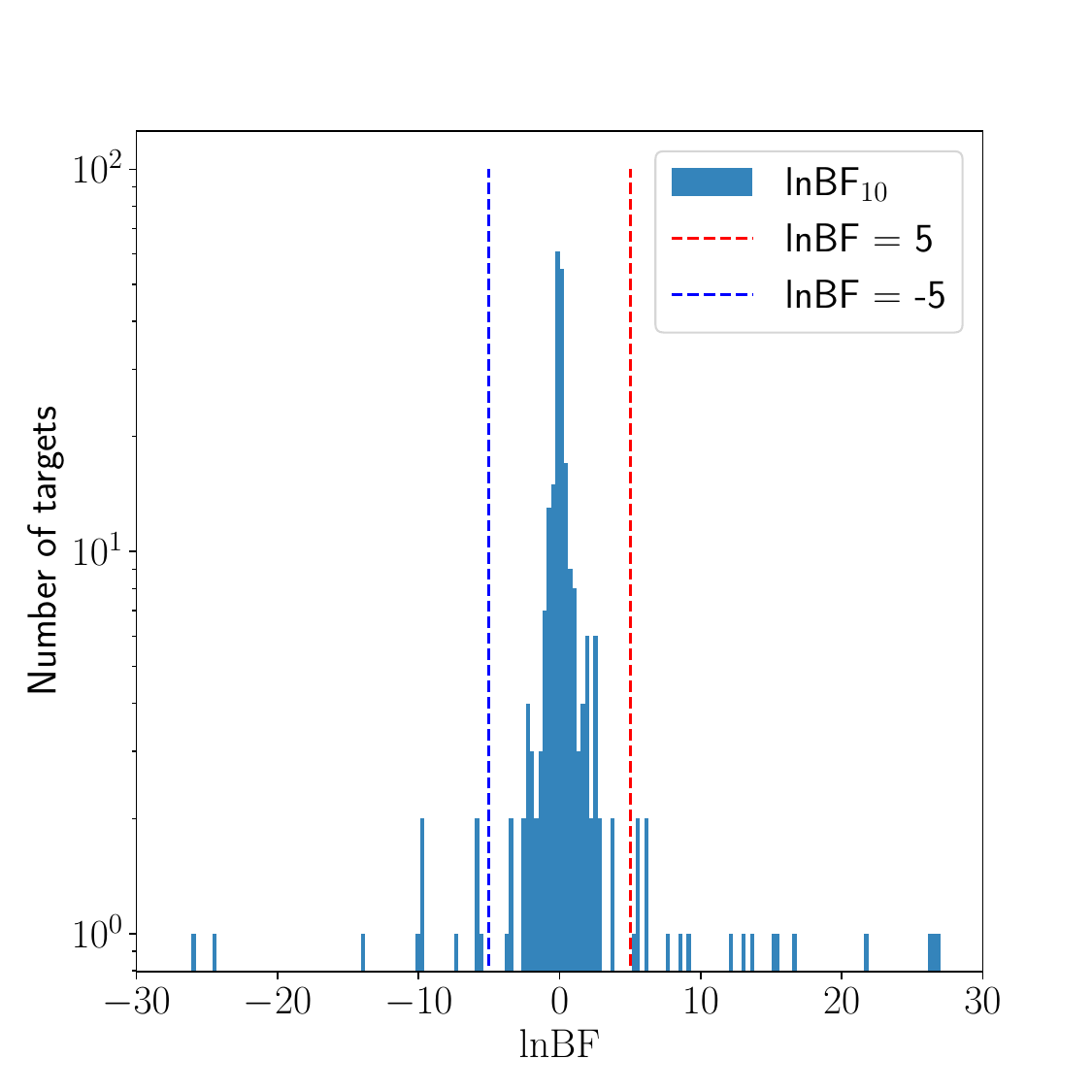}
		\caption{Distribution of $\rm lnBF$ between GDR3PEXO and {GDR2BarCor}. The data with $\rm |lnBF_{10}|>30$ are truncated for better visualization.%; $\rm lnBF_{20}$ is the Bayes factor between OptAstroPEXO model and {\color{red}GDR2BarCor} model; $\rm lnBF_{21}$ is the Bayes factor between OptAstroPEXO model and GDR3PEXO model.%\textbf{Upper panel}: 
} %\textbf{Lower panel}: The dots represent the target stars with both $\rm lnBF_{10}>0$ and $\rm lnBF_{20}>0$; the dashed lines are the thresholds for model selection. 
  
		\label{fig:lnBF10}

	\end{figure}
 
The distribution of $\rm lnBF_{10}$ is shown in Fig. \ref{fig:lnBF10}. The distribution of $\rm |lnBF_{10}|$ for $\rm lnBF_{10}>0$ targets and $\rm lnBF_{10}\le 0$ targets show no significant difference (p-value for K-S test is $\sim 0.19$). This figure indicate that it is reasonable to use $\rm |lnBF_{10}| = 5$ as the threshold for model selection, which is consistent with \citep{Kass1995}.

%%%%%%%%%%%%%%%%%%%%%%%%%%%%%%%%%%%%%%%%%%%%%%%%%%

% Don't change these lines
\bsp	% typesetting comment
\label{lastpage}
\end{document}